\documentclass[trackchanges, twocolumn]{aastex701}
\usepackage{amsmath}
\usepackage{CJKutf8}
\newcommand{\namecn}[1]{\begin{CJK*}{UTF8}{gbsn}({#1})\end{CJK*}}

\begin{document}

\title{HD\,164604\,c: a second giant planet on a 15-yr orbit and the constraint of the planet-planet mutual inclination\footnote{This paper includes data gathered with the 6.5 meter Magellan Telescopes located at Las Campanas Observatory, Chile.}}

\author[orcid=0000-0001-6753-4611]{Guang-Yao Xiao \namecn{肖光耀}}
\affiliation{State Key Laboratory of Dark Matter Physics, Tsung-Dao Lee Institute \& School of Physics and Astronomy, Shanghai Jiao Tong University, Shanghai 201210, China}
\email{gyxiao\_tdli@sjtu.edu.cn}

\author[orcid=0000-0001-6039-0555]{Fabo Feng \namecn{冯发波}}
\affiliation{State Key Laboratory of Dark Matter Physics, Tsung-Dao Lee Institute \& School of Physics and Astronomy, Shanghai Jiao Tong University, Shanghai 201210, China}
\email[show]{ffeng@sjtu.edu.cn}

\author[0000-0003-1305-3761]{R. Paul Butler}
\affiliation{Earth and Planets Laboratory, Carnegie Institution for Science, 5241 Broad Branch Road, NW, Washington, DC 20015, USA}
\email{bluaper@gmail.com}

\author[0009-0008-2801-5040]{Johanna K. Teske}
\affiliation{Earth and Planets Laboratory, Carnegie Institution for Science, 5241 Broad Branch Road, NW, Washington, DC 20015, USA}
\email{jteske@carnegiescience.edu}

\author[0000-0002-8681-6136]{Stephen A. Shectman}
\affiliation{Observatories of the Carnegie Institution for Science, 813 Santa Barbara Street, Pasadena, CA 91101, USA}
\email{shec@carnegiescience.edu}

\author[0000-0002-5226-787X]{Jeffrey D. Crane}
\affiliation{Observatories of the Carnegie Institution for Science, 813 Santa Barbara Street, Pasadena, CA 91101, USA}
\email{crane@carnegiescience.edu}

\author[0000-0002-6937-9034]{Sharon X. Wang \namecn{王雪凇}}
\affiliation{Department of Astronomy, Tsinghua University, Beijing 100084, People’s Republic of China}
\email{sharonw@mail.tsinghua.edu.cn}








\begin{abstract}

We report the discovery of a new massive giant planet, HD\,164604\,c ($a_c = 5.556_{-0.10}^{+0.093}$\,au, $e_c = 0.196_{-0.078}^{+0.078}$ and $m_c = 9.5_{-1.25}^{+1.2}$ or $7.6_{-1.0}^{+1.0}\,M_{\rm Jup}$), orbiting a K3.5 dwarf, The result is based on the combined analysis of high-precision radial-velocity data, Hipparcos, and Gaia DR2 and DR3 astrometry. We refine the orbital parameters of the inner planet HD\,164604\,b to $a_b = 1.362_{-0.012}^{+0.012}$\,au, $e_b = 0.479_{-0.021}^{+0.027}$, and $m_b = 13.2_{-1.5}^{+1.8}\,M_{\rm Jup}$ (or $8.8_{-1.5}^{+1.9}\,M_{\rm Jup}$). 
Depending on the two possible orbital orientations of HD\,164604\,c, the true mutual inclination between the two planets is $\psi_{bc}=5.0^{+3.7}_{-2.2}$$^\circ$ (prograde) or $162.1^{+7.1}_{-4.7}$$^\circ$ (retrograde).
Long-term N-body integrations show that most orbits with the retrograde configuration remain dynamically stable for at least 10 Myr, while orbits with the prograde motion might rapidly evolve into chaos or lead to ejection. 
The retrograde architecture points to a violent dynamical history, possibly involving von Zeipel–Lidov–Kozai cycles or three-body scattering, while the prograde scenario might be consistent with coplanar and mild disk migration. 
Future Gaia DR4 astrometry will break the inclination degeneracy and distinguish between prograde and retrograde orbits for HD\,164604\,c.

\end{abstract}

\keywords{\uat{Stellar astronomy}{1583}}

\section{Introduction}

Since the first exoplanet orbiting a solar-type star — 51\,Pegasi\,b — was detected via the radial-velocity (RV) technique in 1995 \citep{Mayor1995}, exoplanet science has flourished. To date, over 6000 exoplanets have been discovered and confirmed through RV, transit, direct imaging, astrometry, and microlensing \citep{Christiansen2025PSJ}.

The mutual inclination ($\psi$) between planets is a key probe of the dynamical history of multi-planetary systems, yet it remains challenging to measure directly, especially in non-transiting systems due to limited information on orbital orientation. To date, only a handful of systems possess well-constrained true mutual inclination. Roughly half of them are determined via dynamical modeling of transit timing and/or duration variations (TTVs and TDVs; e.g., \citealt{Masuda2017AJ, Maciejewski2025}). The remains are obtained from the joint analysis of multiple techniques, including RV, Hipparcos-Gaia astrometry (e.g.,\citealt{Brandt2018,Brandt2021}), and ground- or space-based relative astrometry (e.g., HST and JWST). 

Recent studies have revealed several systems with relatively significant misalignment between planets. One notable example is HAT-P-11, which consists of a transiting super-Neptune planet on a $\sim5$-day orbit \citep{Bakos2010ApJ} and a non-transiting super-Jupiter on a $\sim10$-year orbit \citep{Yee2018AJ}. \citet{An2025AJ} presented the first observational evidence that the outer planet is significantly misaligned with both the host star and and the inner transiting planet. \citet{Lu2025ApJ} proposed that the present-day configuration of this system can be reproduced by planet–planet scattering \citep{Ford2008} followed by von-Zeipel–Lidov–Kozai (ZLK; \citealt{vonZeipel1910AN,Lidov1962,Kozai1962}) cycles and high-eccentricity migration. Because the inner planet’s astrometric signature of such systems is negligible, the astrometric component of the orbital fit is well approximated by a single-planet model, and thus allows the inclination of the outer planet to be measured.

However, in non-transiting multi-planet systems, the degeneracy of orbital orientation becomes even severe \citep{Kervella2020A&A,Feng2025MNRAS}, because the inclinations of all planets must be constrained.
For example, by jointly analysing RV data, Hipparcos-Gaia absolute astrometry (also known as proper motion anomaly, PMA) and JWST relative astrometry, \citet{Bardalez2025ApJ} recently reported two possible values of $\psi_{bc}={32_{-15.1}^{+13.6}}^{\circ}$ and $\psi_{bc}={145.0_{-11.1}^{+15.8}}^{\circ}$ for the non-transiting 14\,Her system. Using additional astrometry from Gaia second data release (DR2), \citet{Xiao2025RNAAS} further constrained the orbital orientation of the inner planet, 14\,Her\,b, and found a definitive true mutual inclination of $\psi_{bc}={35.3_{-7.3}^{+6.8}}^{\circ}$, consistent within $2\sigma$ with the value of $62^{\circ}\pm12^{\circ}$ obtained by \citet{Benedict2023AJ}. 

In this study, we combine RVs, astrometry from Hipparcos and Gaia second and third data releases (GDR2 and GDR3; \citealt{GaiaCollaboration2018,GaiaCollaboration2023}), and the prior inclination information from Gaia two-body solution \citep{Gaia_two_body_2023} to simultaneously determine the orbits of both non-transiting planets in the HD\,164604 system, This system hosts a previously known giant planet on an eccentric $\sim650$-day orbit \citep{Arriagada2010ApJ}. we present the discovery of HD\,164604\,c, a second giant planet on a $\sim5400$\,day orbit exterior to HD\,164604\,b. We first apply this joint analysis to the relatively short-period planet ($<1000$\,days), a regime that previous works have largely neglected.  We further characterized the three-dimensional (3D) orbit and the true mutual inclination of the two planets. Compared with existing approaches (e.g., \citealt{Xuan2020MNRAS,Brandt2021borvara,Venner2021,Kervella2022}), a key improvement of our method is the incorporation of GDR2, which allows for constraints on systems with periods shorter than satellite observational baseline ($\sim1000$\,days).

This paper is organised as follows. Section~\ref{sec:obs} describes the spectroscopic and astrometric observations of HD\,164604. In Section~\ref{sec:stellar}, we derive the fundamental stellar properties. Section~\ref{sec:orb} presents the orbital solution and planetary parameters. Section~\ref{sec:dis} discusses orbital stability and planetary formation scenarios. A brief summary is provided in Section~\ref{sec:con}.

\section{Observations}
\label{sec:obs}
\subsection{Radial Velocities}
\subsubsection{MIKE}
The first spectrum of HD\,164604 was obtained in 2003 June using the MIKE echelle spectrograph \citep{Bernstein2003}, mounted on the 6.5 m Magellan II (Clay) telescope located at Las Campanas Observatory in Chile. An iodine absorption cell \citep{Marcy1992PASP} is mounted in front of the MIKE entrance slit, providing a wavelength reference in the range of $5000\sim6200$ \AA. The slit width was set to 0.\arcsec35, corresponding to a resolution of $R=\lambda/\Delta\lambda\sim50,000$. With the MIKE spectrograph, \citet{Arriagada2010ApJ} obtained 18 precise RV observations spanning 6 years with a median uncertainty of $3.44\,{\rm m\,s^{-1}}$, leading to the discovery of the cold Jupiter, HD\,164604\,b ($a_b=1.3\pm0.05\,\rm au$, $e_b=0.24\pm0.14$ and $M_b{\rm sin}i=2.7\pm1.3\,M_{\rm Jup}$), on a moderately eccentric orbit. We include their RVs in our analysis.
\subsubsection{PFS}
The Carnegie Planet Finder Spectrograph (PFS; \citealt{Crane2006, Crane2008, Crane2010}) on the 6.5 m Magellan II Telescope was designed for high-resolution spectroscopic measurements. It spans a wavelength range of 391–734 nm at a resolving power of R $\sim$ 130,000 through a 0.\arcsec3$\times$2.\arcsec5 slit. As with MIKE, an iodine cell was utilized for wavelength calibration. Typically, PFS achieves an RV precision of $0.5-1.0$ ${\rm m\,s^{-1}}$ for nearby, bright, and Chromospherically quiet stars (e.g., $d\lesssim$100\,pc, $V$$\approx8-10$\,mag, and $\log R'_{\text{HK}}\approx5$, \citealt{Arriagada_2011}). The process of spectral reduction and RV extraction was carried out through the custom pipeline of \citet{Butler_1996}. 
Two chromospheric activity indices are also generated automatically. The $S$ index is derived from the emission reversal at the cores of the Fraunhofer H and K lines of Ca II at 3968\AA\,and 3934\AA, respectively. The $H$ index quantifies the chromospheric emission component at the $H_\alpha$ Balmer line of hydrogen at 6563\AA\citep{daSilva2011A&A}. Details can be found in \citet{Butler2017AJ}.
Since 2010, targets from the Magellan Planet Search Program have been moved from MIKE to PFS for improved precision and long-term stability. Using PFS, we collected 35 precise RVs (median uncertainty of $1.06\,{\rm m\,s^{-1}}$) from April 2010 to June 2024, extending the total RV baseline of HD\,164604 to about 21 years. RV data used in this work are presented in the Appendix~\ref{APPE_A}.

\subsection{Hipparcos and Gaia Astrometry}
To derive astrometric constraints, we incorporate the absolute astrometry (positions in R.A. $\alpha$, and decl. $\delta$, proper motions $\mu_{\alpha}$ and $\mu_{\delta}$, and parallax $\varpi$) of the new Hipparcos reduction \citep{vanLeeuwen2007} and Gaia second and third data releases (GDR2 and GDR3; \citealt{GaiaCollaboration2018,GaiaCollaboration2023}), Hipparcos intermediate astrometry data (IAD; also refer to as abscissae data, \citealt{vanLeeuwen2007}), and synthetic epoch data from Gaia Observation Forecast Tool \footnote{\url{https://gaia.esac.esa.int/gost/index.jsp}} (GOST) to perform joint analysis with RVs\footnote{Data available on \dataset[Zenodo]{https://doi.org/10.5281/zenodo.18172547}}. The Hipparcos IAD and Gaia GOST data mainly comprise the scan angle $\psi$ of the satellite, the along-scan (AL) parallax factor $f^{AL}$, and associated observation epoch at barycenter. Since the Gaia epoch astrometry is not available, we use GOST to predict the Gaia observations. The choice of Hipparcos version has negligible impact on our analyses, because we directly model the systematics in Hipparcos IAD using offsets and jitters for a given target \citep{Feng2023MNRAS}. Overall, the astrometric anomaly between two satellites ($\rm \sim25\,yr$) primarily constrains the outer long-period companion, whereas the shorter time-scale anomaly between GDR2 and GDR3 is used to determine the inner one. 

According to the Hipparcos-Gaia catalog of astrometric accelerations (HGCA; \citealt{Brandt2018, Brandt2021}), HD\,164604 has a $\chi^2$ value of 53.8 ($6.9\sigma$ detections assuming Gaussian errors and two degrees of freedom) under the hypothesis of constant proper motion, indicating a significant astrometric signature induced by the outer companion. 
Unlike HGCA-based approaches, we adopt a case-by-case strategy that models astrometric systematics a posteriori using jitter and offsets terms. This method has proven effective in preventing the inflation of uncertainties during frame transformations \citep{Feng2021MNRAS}. 

HD\,164604 also appears in the Gaia DR3 non-single-star catalog and its two-body solutions, such as orbital period, eccentricity and Thiele-Innes (TI) coefficients, are already provided. Gaia-only measurements confirm that HD\,164604\,b is a super-Jupiter with a mass of $14.3\pm5.5\,M_{\rm Jup}$ and an inclination of $29^\circ\pm19^\circ$ \citep{Gaia_two_body_2023}. Although some studies have applied Gaia-only solutions (mainly TI elements) along with RVs to further constrain the planet's orbit (e.g., \citealt{Winn2022AJ,Stevenson2023MNRAS,Stefansson2025AJ}), we do not include them in our model due to the multiplicity of HD\,164604 system. Instead, we only utilize its inclination measurement as a priori in our sampling. This informative prior is helpful to limit the orbital orientation of the inner planet, HD\,164604\,b.

\section{Stellar Characterization}
\label{sec:stellar}
\subsection{Spectral Energy Distribution}

\begin{figure}[ht!]
    \centering
    \includegraphics[width=0.48\textwidth]{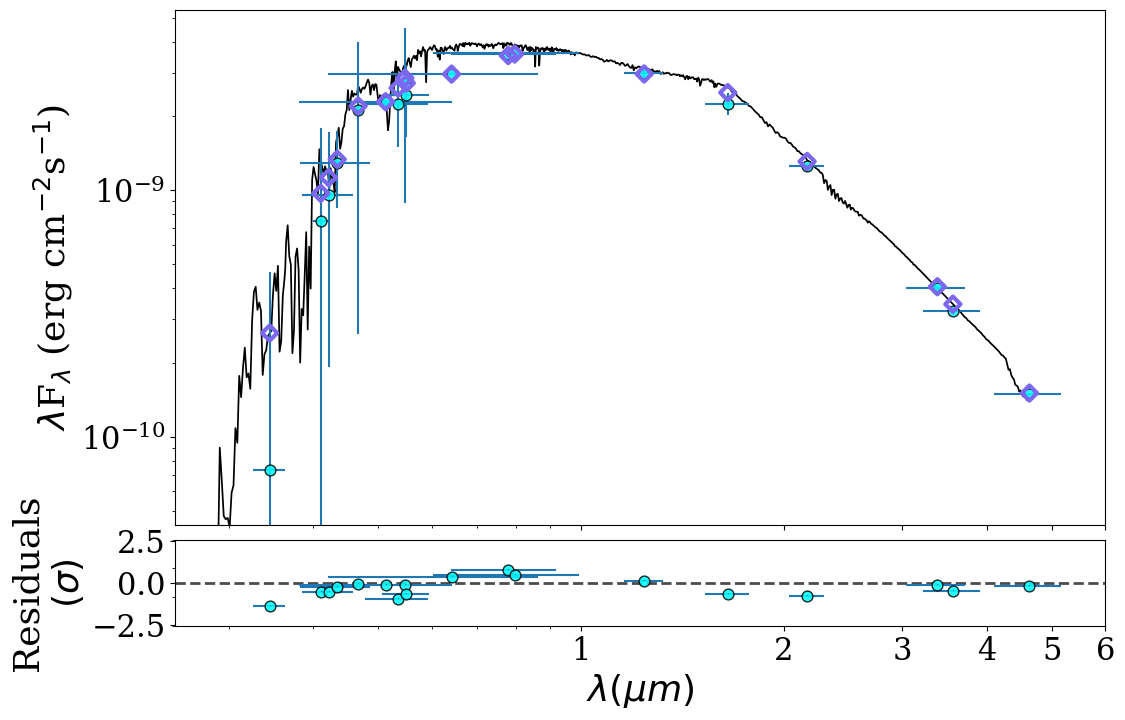}\caption{\textbf{Top panel:} Best-fit SED to archival photometry of HD\,164604. Circles mark the observed photometric points, while diamonds indicate the corresponding synthetic fluxes generated by the model. Horizontal bars denote the effective width of each filter. \textbf{Bottom panel:} Residuals between the observations and the model, normalized to the photometry uncertainties. 
\label{fig:sed}}
\end{figure}

\begin{figure}[ht!]
    \centering
    \includegraphics[width=0.40\textwidth]{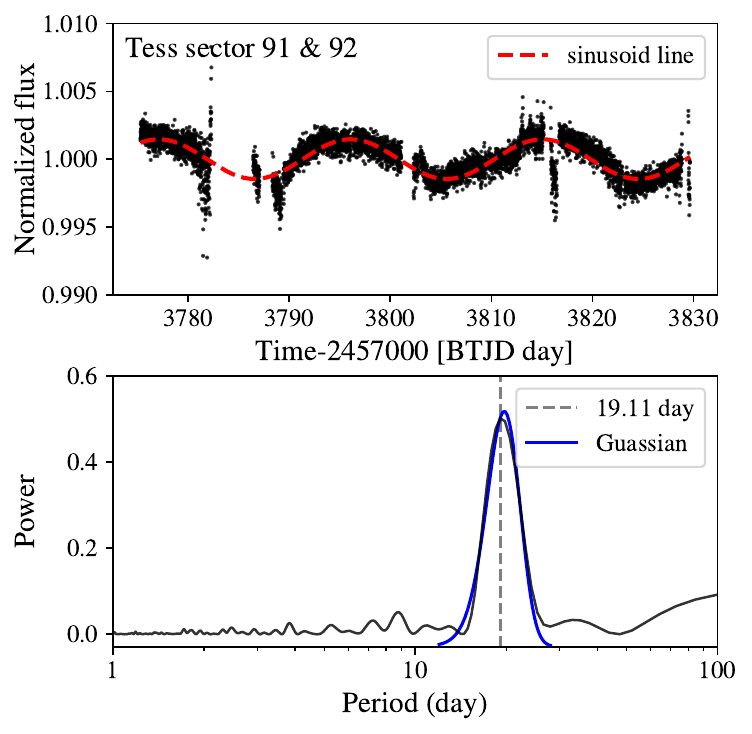}\caption{\textbf{Top panel:} Detrended TESS light curves of HD\,164604 along with a sinusoid model (red dashed line) at $P=19.11$\,days. \textbf{Bottom panel:} the Lomb-Scargle periodogram of the light curves with a peak at 19.11\,days. A Gaussian model (blue line) is fitted to the peak to obtain the uncertainty of the rotation period. 
\label{fig:prot}}
\end{figure}

\begin{table}
\caption{Stellar Parameters}
\label{tab:stellar}
\begin{tabular}{lcc}
\hline\hline
Parameter & Value & Source \\
\hline
\multicolumn{3}{c}{Basic Properties}\\
HIP ID & 88414& 1\\
Gaia ID & DR3 4062446910648807168& 2\\
TIC &  406317882 & 3\\
TESS Mag. & 8.6834 & 3\\
Sp. Type & K3.5V & 6\\
$\varpi$ (mas)& $25.376\pm0.062$ &2\\
$V$ (mag) & $9.83\pm0.04$ & 4\\
$B - V$ & $1.05\pm0.08$ & 4 \\
\multicolumn{3}{c}{Spectroscopy}\\
$T_{\rm eff}\,(\rm K)$ &$4684\pm157$& 5\\
${\rm log}\,\textsl{g}$ (cgs) &$4.32\pm0.41$& 5\\
$\rm [Fe/H]$ (dex) &$0.12\pm0.07$&5\\
\multicolumn{3}{c}{SED \& Isochrones}\\
$T_{\rm eff}\,(\rm K)$ &$4663\pm31$& 6\\
${\rm log}\,\textsl{g}$ (cgs) &$4.49\pm0.41$& 6\\
$\rm [Fe/H]$ (dex) &$0.12\pm0.07$& 6\\
$L_{\star}\,(L_{\sun})$ &$0.26\pm0.01$& 6\\
$M_{\star}\,(M_{\sun})$  & $0.77\pm0.02$ & 6\\
$R_{\star}\,(R_{\sun})$  & $0.78\pm0.02$ & 6\\
${\rm Age} $ (Gyr)  &$10.7\pm3.8$ & 6\\
\multicolumn{3}{c}{Photometry}\\
$P_{\rm rot} $ (days)  &$19\pm3$ & 6\\
\multicolumn{3}{c}{Gyrochronology}\\
${\rm Age} $ (Gyr)  &$1.6\pm0.7$ & 6\\
\hline
\end{tabular}
\tablerefs{(1) \citet{ESA1997}, (2) \citet{GaiaCollaboration2023}, (3) \citet{Stassun2018}, (4) \citet{Perryman1997},
(5) \citet{Santos2013}, (6) this work}
\end{table}

To determine the stellar properties of HD\,164604, we conduct Spectral Energy Distribution (SED) fitting with the publicly available \texttt{astroARIADNE} package \citep{Vines_2022}. 
This code can automatically retrieve the Gaia distance and available broad-band photometry from multiple archives, including GALEX FUV, Tycho-2 BT/VT, 2MASS JHKs, WISE W1–W2, TESS, and Gaia G, RP, and BP magnitudes. To measure the effective temperature $T_{\rm eff}$, Fe abundance ${\rm [Fe/H]}$, surface gravity ${\rm log}\,\textsl{g}$, extinction ${A_V}$ and radius ${R_\star}$ of a star, \texttt{astroARIADNE} employs Bayesian Model Averaging to incorporate all stellar atmospheric models and the MIST isochrones \citep{Paxton2011, Choi2016, Dotter2016}. 

In our case, we restrict the model grid to the PHOENIX \citep{Husser2013A&A} and Castelli \& Kurucz \citep{Castelli2003} atmospheric models, and adopt Gaussian priors on $T_{\rm eff}$, ${\rm [Fe/H]}$ and ${\rm log}\,\textsl{g}$ using the spectroscopic analysis of \cite{Santos2013}. Given the proximity of the star, we fix the extinction ${A_V}$ to zero. 
The best-fit results suggest that HD\,164604 is a K3.5V main-sequence star with a mass of $0.77\pm0.02\,M_{\odot}$, which is well consistent with the estimate of \citet{Santos2013}, $0.77\pm0.14\,M_{\odot}$. However, the age is loosely constrained. We thus use \texttt{stardate}\citep{Angus2019AJ}, which combines isochrone fitting with gyrochronology, to redetermine this quantity.
Figure \ref{fig:sed} shows the resulting plot of SED fitting and Table~\ref{tab:stellar} lists the detailed stellar parameters.

\subsection{Stellar Rotation Period}
To determine the rotation period of HD\,164604, we accessed all available TESS Target Pixel File data \citep{Ricker2015} and then conducted a simple aperture photometry to extract light curves. We use the open-source systematics insensitive periodogram (SIP; \citealt{Hedges2020RNAAS}) package for this purpose. 
Figure~\ref{fig:prot} presents the detrended light curves of HD\,164604 at a 2-minutes cadence for Sector 91 \& 92, along with the periodogram of the stitched light curves. We derived a rotation period of $P_{\rm rot}=19\pm3$\,days. The uncertainty includes contributions from the width (Gaussian fit) of the periodogram peak and an additional $10\%$ to account for stellar surface differential rotation \citep{Epstein2014ApJ, Claytor2022ApJ}.

\section{Orbital analysis and result}
\label{sec:orb}
\subsection{Periodogram}
\begin{figure}[ht!]
    \centering
    \includegraphics[width=0.48\textwidth]{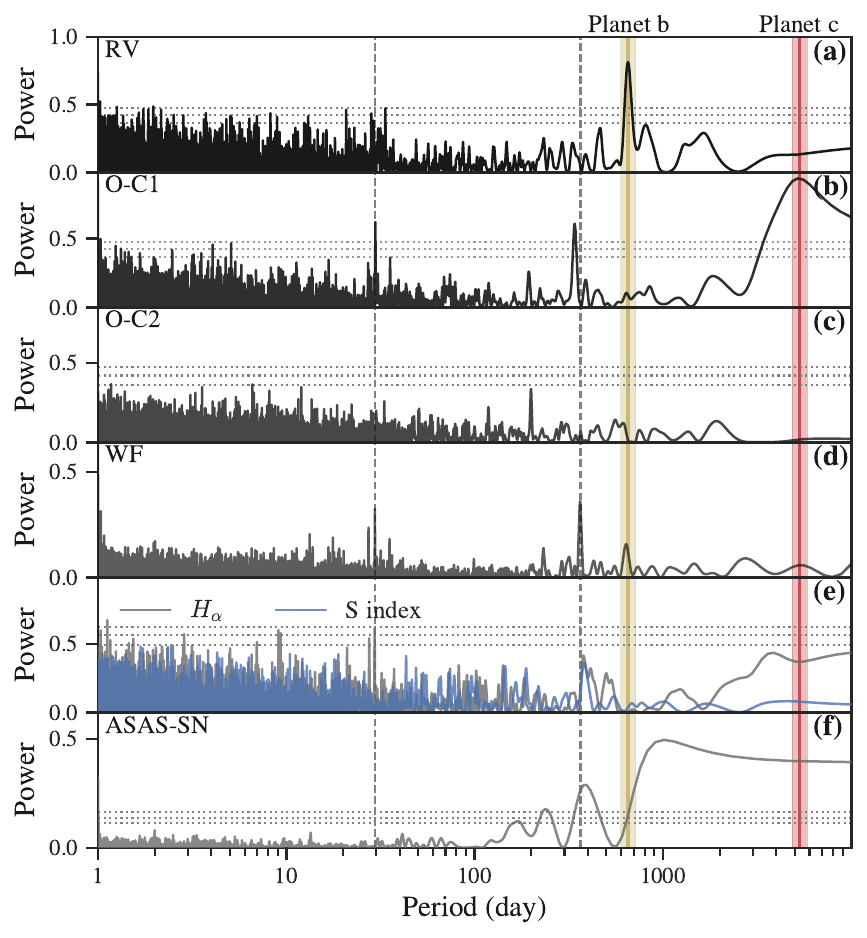}\caption{Generalized Lomb-Scargle (GLS) periodograms for HD\,164604. \textbf{Panel (a):} the GLS periodograms of all the observed RVs. \textbf{Panel (b):} the residual RVs after subtracting the signal of the inner planet, HD\,164604\,b. \textbf{Panel (c):} the residuals after subtracting both planet signals. \textbf{Panel (d):} window function. \textbf{Panel (e):} stellar activity indices extracted from PFS spectra. \textbf{Panel (f):} photometry of ASAS-SN. The signal of planet b and c are respectively denoted with yellow and red shade regions. The horizontal grey lines in each panel, top to bottom, indicate the 0.001, 0.01, 0.1 False Alarm Probability (FAP) levels. A monthly (30\,days) and yearly (365\,days) signals are denoted with vertical dashed lines. 
\label{fig:periodogram}}
\end{figure}
Figure~\ref{fig:periodogram} shows the Generalized Lomb-Scargle Periodogram \citep{Zechmeister2009} for HD\,164604 RVs, RV residuals, stellar activity indices (i.e., $S$ index and $H_\alpha$ value extracted from PFS spectra), and light curve from ASAS-SN. In panel (a), the period peak of 650 days corresponds to the signal induced by the inner planet, HD\,164604\,b. After subtracting this signal from the RVs, a longer signal near 5400 days emerges (panel b). 
Because the activity indices are only available for PFS and span only 5058 days, which are slightly shorter than the outer signal, the periodogram shows no significant peaks beyond 5000 days.
We further calculate Pearson’s correlation coefficient $r$ of the $S$ index and $H_\alpha$ with RVs and residuals, and find that there is no strong correlation between stellar activities and observed RVs (see Appendix~\ref{APPE_B}).
We therefore adopt a white-noise model rather than a red-noise model for the RV modeling.

\subsection{Gaia proper motion anomaly}
\begin{figure}[ht!]
    \centering
    \includegraphics[width=0.45\textwidth]{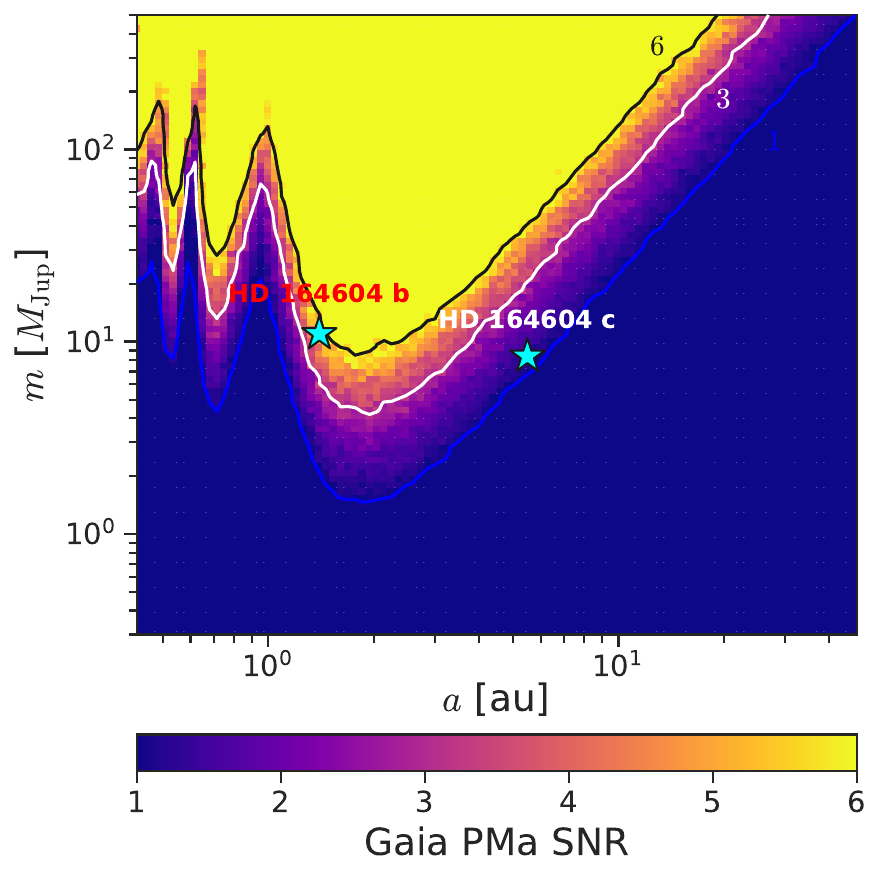}\caption{Gaia PMa SNR map across the $m-a$ space. The blue, white and black contour lines respectively correspond to SNR=$1$, $3$ and $6$. The two cyan stars denote the currently measured positions of two planets in $m-a$ space. It is evident that the PMa of HD\,164604 between GDR2 and GDR3 is dominated by the inner planet. We find the simulated SNR of HD\,164604\,b is within $3\sim6$, slightly lower than the observed value of $6.7$. This discrepancy is attributed to the uniform sampling of its inclination. If the inclination is fixed to a lower value ($<20^\circ$), corresponding to a nearly face-on configuration, the contour line will shift towards a lower mass and could align well with the observation. In other words, the SNR map presented here reflects only the average behaviour and may differ from any one specific orbital configuration.   
\label{fig:Gaia_pma}}
\end{figure}

To quantify the proper motion discrepancy of HD\,164604 between GDR2 and GDR3 induced by its planets, we simulate Gaia epoch data with GOST across a range of orbital configurations, fit each release separately with a four-parameter astrometric model ($\varpi$ is fixed) to derive their synthetic proper motions (for detailed models, see Equation C26 and C27 of \citealt{Xiao2024MNRAS}), and then compute the PMa signal-to-noise ratio (SNR). The semi-major axis $a$ is uniformly sampled from $0.4\sim34$\,au, and the companion mass $m$ from $0.3\sim500\,M_{\rm Jup}$ ($100\times100$ grids with 25 orbits for each grid).
All remaining orbital elements, except eccentricity, which is fixed at 0, are drawn from the same uniform distributions as their priors (see Table~\ref{Tab:result}). 
We only consider a single-planet model in our simulation.
We define the SNR of PMa between two catalogs as
\begin{equation}
    {\rm SNR} = \sqrt{(\Delta \mu)\,\Sigma^{-1}\,(\Delta \mu)^T},
\end{equation}
where $\Delta \mu=[\Delta\mu_\alpha, \Delta\mu_\delta]$ is the vector of the proper motion difference and $\Sigma$ is the relevant covariance matrix. For each grid, we calculate the mean SNR of 25 orbits.
Figure~\ref{fig:Gaia_pma} shows the PMa SNR map assuming $M_\star=0.77\,M_\odot$, $\varpi=24.99$\,mas.
According to the Gaia archive, we find $\Delta\mu_{\alpha}=0.10\pm0.11\,{\rm mas\,yr^{-1}}$ and $\Delta\mu_{\delta}=0.57\pm0.09\,{\rm mas\,yr^{-1}}$ between GDR2 and GDR3, respectively, which correspond to a PMa SNR of 6.7. 
The high SNR largely aligns with our simulations, which could support the scenario that the inner planet HD\,164604\,b is massive enough to drive the PMa observed between GDR2 and GDR3. Therefore, the implementation of modeling multiple Gaia data releases enables the constraint of systems with orbital periods shorter than the observational baseline.    

\subsection{Joint modeling of RV and Hipparcos-Gaia astrometry}
We adopt a right-handed coordinate system defined by the orthonormal triad $[\hat{x}, \hat{y}, \hat{z}]$, where $\hat{x}$ points the direction of increasing R.A., $\hat{y}$ points the direction of increasing decl., and $\hat{z}$ points away from the observer (convention $\mathrm{III}$ of \citealt{Feng2019ApJS..242...25F}). 
This convention is used in the astrometry models of Hipparcos and Gaia \citep{ESA1997, Lindegren2012A&A}. 

We simultaneously constrain the 3D orbit of both the inner and outer planets through a joint fit to the RVs and Hipparcos-Gaia astrometry, following the method developed by \citet{Feng2023MNRAS} and modified by \citet{Xiao2024MNRAS}. 
The primary fitted parameters involved in our model include seven elements (orbital period $P$, RV semi-amplitude $K$, eccentricity $e$, argument of periastron $\omega$ of stellar reflex motion, orbital inclination $i$, longitude of ascending node $\Omega$, mean anomaly $M_{0}$ at the minimum epoch of RV data), five astrometric offsets ($\Delta \alpha*$, $\Delta \delta$, $\Delta \varpi$, $\Delta \mu_{\alpha*}$ and $\Delta \mu_\delta$) of barycenter relative to GDR3, RV zero-points $\gamma$ and jitter terms $\sigma_{\rm jit}$ for MIKE and PFS, astrometric jitter $J_{\rm hip}$ for Hipparcos IAD, and error inflation factor $S_{\rm gaia}$ for Gaia astrometry. The semi-major axis $a$ of the planet relative to the host, the mass of planet $m_{\rm p}$, and the epoch of periastron passage $T_{\rm p}$ can be derived from above orbital elements.

For RV model, the likelihood can be calculated by
\begin{equation}
\begin{split}
  \mathcal{L}_{\rm RV}
  &=\prod\limits_{j=1}^{N_{\rm RV}}\prod\limits_{k=1}^{N_{\rm inst}}
    \frac{1}{\sqrt{2\pi(\sigma_{j,k}^2+\sigma_{{\rm jit},k}^2)}}\times\\
  &\quad\exp\!\left(-\frac{(v_{j,k}-\hat{v}_{j,k}-\gamma_k)^2}
    {2(\sigma_{j,k}^2+\sigma_{{\rm jit},k}^2)}\right)~,
\end{split}
\end{equation}
where $N_{\rm RV}$ and $N_{\rm inst}$ are the number of RV measurements and instruments, and $\gamma_k$ and $\sigma_{{\rm jit},k}$ are the RV zero-point and the jitter term. $\hat{v}_{j,k}$ and $v_{j,k}$ are the predicted and observed RVs, respectively.

The likelihood for Hipparcos IAD is expressed as
\begin{equation}
  \mathcal{L}_{\rm hip}=\prod\limits_{j=1}^{N_{\rm IAD}}\frac{1}{\sqrt{2\pi(\sigma_{j}^2+J_{\rm hip}^2)}}\times{\rm exp}\left(-\frac{(\hat{\xi}_j-\xi_j)^2}{2(\sigma_{j}^2+J_{\rm hip}^2)}    \right)~,
\end{equation}
where $N_{\rm IAD}$ is the total number of Hipparcos IAD, $\sigma_j$ is the individual measurement uncertainty, and $J_{\rm hip}$ is the jitter term. Predicted and observed 1D abscissae data are denoted with $\hat{\xi}_j$ and $\xi_j$, respectively. The abscissae $\hat{\xi}_j$ is synthesized by projecting the combined motion of the system's barycenter, the stellar reflex motion, and the parallax motion onto the 1D scan direction of Hipparcos. In addition, the offset between Hippacos astrometry and the astrometry propagated from the GDR3 epoch to the Hipparcos epoch are also incorporated into the modeling of $\hat{\xi}_j$.

The likelihood for GDR2 and GDR3 can be written as
\begin{equation}
\begin{split}
  \mathcal{L}_{\rm gaia}=
  &\prod\limits_{k=1}^{N_{\rm DR}}\frac{1}{\sqrt{(2\pi)^5|\Sigma_k(S^2)|}}\times\\
  &{\rm exp}\left(-\frac{1}{2}(\Delta\hat{\vec{\iota}}_k-\Delta\vec{\iota}_k)^{T}[\Sigma_k(S^2)]^{-1}(\Delta\hat{\vec{\iota}}_k-\Delta\vec{\iota}_k)    \right)~,
\end{split}
\end{equation}
where $N_{\rm DR}$ represents the number of Gaia data releases ($N_{\rm DR}=2$ when we use both GDR2 and GDR3), $\Sigma_k$ is the covariance provided by Gaia catalog, and $S$ is the error inflation factor. $\Delta\vec{\iota}_k$ is the five-parameter astrometric vector for catalog $k$ expressed relative to GDR3, e.g.,
\begin{equation}
\begin{split}
\Delta\vec{\iota}_k\equiv
&\,((\alpha_k-\alpha_{\rm DR3})\,{\rm cos\delta_k},\,\delta_k-\delta_{\rm DR3},\,\\
&\varpi_k-\varpi_{\rm DR3},\,\mu_{\alpha k}-\mu_{\alpha \rm DR3},\,\mu_{\delta k}-\mu_{\delta \rm DR3}),
\end{split}
\end{equation}
whereas $\Delta\hat{\vec{\iota}}_k$ is the five synthetic astrometric vector  obtained by fitting the standard five-parameter astrometric model to the simulated abscissae. More details can be found in Appendix B of \citet{Xiao2024MNRAS}.

\begin{table*}[!]
\centering
\caption{Parameters for HD\,164604 system$^*$}\label{Tab:result}
\begin{tabular*}{\textwidth}{@{}@{\extracolsep{\fill}}lccccc@{}}
\hline \hline
Parameter & Prograde & Retrograde & \multicolumn{2}{c}{MAP value$\rm ^a$} & Prior${\rm ^b}$ \\
&$i_c<90^\circ$&$i_c>90^\circ$&$i_c<90^\circ$&$i_c>90^\circ$&\\
\hline
\textbf{HD\,164604\,b}&\\
Orbital period $P_b$ (days)&\multicolumn{2}{c}{${653.9}_{-1.7}^{+1.5}$}&653.8&655.2&Log-$\mathcal{U}(-1,16)$\\
RV semi-amplitude $K_b$ (m s$^{-1}$)&\multicolumn{2}{c}{${76.2}_{-1.8}^{+2.3}$}&75.1&77.5&$\mathcal{U}(10^{-6},10^{6})$\\
Eccentricity $e_b$&\multicolumn{2}{c}{${0.479}_{-0.021}^{+0.027}$}&0.460&0.511&$\mathcal{U}(0,1)$\\
Argument of periapsis $\omega_b$ ($^\circ$)$\rm ^c$&\multicolumn{2}{c}{${64.6}_{-2.0}^{+1.7}$}&64.1&66.0&$\mathcal{U}(0,360)$\\
Mean anomaly $M_{0,\,b}$ ($^\circ$)&\multicolumn{2}{c}{${157.0}_{-5.3}^{+4.9}$}&157.2&160.5&$\mathcal{U}(0,360)$\\
Longitude of ascending node $\Omega_b$ ($^\circ$)&\multicolumn{2}{c}{ ${123}_{-26}^{+25}$}&129&144&$\mathcal{U}(0,360)$\\
Inclination $i_b$ ($^\circ$)&${10.6}_{-1.3}^{+1.4}$&${16.0}_{-2.8}^{+3.5}$&11.5&16.8&$\mathcal{N}(29,19)$\\
\hline
\textit{Derived parameters}\\
Orbital period $P_b$ (yr)& \multicolumn{2}{c}{${1.7903}_{-0.0047}^{+0.0042}$}&1.7900&1.7939&---\\
Semi-major axis $a_b$ (au)&\multicolumn{2}{c}{${1.362}_{-0.012}^{+0.012}$}&1.353&1.367&---\\
Periapsis epoch $T_{\rm p,b}-2400000$ (JD)&\multicolumn{2}{c}{${53831.3}_{-6.6}^{+7.1}$}&53830.9&53826.9&---\\
Companion mass $m_{b}$ ($M_{\rm Jup}$)&${13.2}_{-1.5}^{+1.8}$&${8.8}_{-1.5}^{+1.9}$&11.9&8.4&---\\
\hline
\textbf{HD\,164604\,c}&\\
Orbital period $P_c$ (days)&\multicolumn{2}{c}{${5387}_{-127}^{+120}$}&5353&5426&Log-$\mathcal{U}(-1,16)$\\
RV semi-amplitude $K_c$ (m s$^{-1}$)&\multicolumn{2}{c}{${19.9}_{-1.0}^{+1.1}$}&19.8&20.3&$\mathcal{U}(10^{-6},10^{6})$\\
Eccentricity $e_c$&\multicolumn{2}{c}{${0.196}_{-0.078}^{+0.078}$}&0.152&0.296&$\mathcal{U}(0,1)$\\
Argument of periapsis $\omega_c$ ($^\circ$)&\multicolumn{2}{c}{${89}_{-10}^{+12}$}&76&84&$\mathcal{U}(0,360)$\\
Mean anomaly $M_{0,\,c}$ ($^\circ$)&\multicolumn{2}{c}{${350.7}_{-12}^{+6.7}$}&359&353&$\mathcal{U}(0,360)$\\
Longitude of ascending node $\Omega_c$ ($^\circ$)&${143.1}_{-10}^{+9.4}$&${18.9}_{-9.3}^{+10}$&140.5&6.8&$\mathcal{U}(0,360)$\\
Inclination $i_c$ ($^\circ$)&${8.6}_{-1.0}^{+1.3}$&${169.3}_{-1.7}^{+1.3}$&8.9&169& Cos$i$-$\mathcal{U}(-1,1)$\\
\hline
\textit{Derived parameters}\\
Orbital period $P_c$ (yr)&\multicolumn{2}{c}{${14.75}_{-0.35}^{+0.33}$}&14.66&14.85&---\\
Semi-major axis $a_c$ (au)&\multicolumn{2}{c}{${5.556}_{-0.10}^{+0.093}$}&5.496&5.596&---\\
Periapsis epoch $T_{\rm p,c}-2400000$ (JD)&\multicolumn{2}{c}{${52949}_{-99}^{+173}$}&52819.1&52908.8&---\\
Companion mass $m_{c}$ ($M_{\rm Jup}$)&${9.5}_{-1.2}^{+1.2}$&${7.6}_{-1.0}^{+1.0}$&9.0&7.7&---\\
\hline
\textbf{Barycenter parameters}\\
$\alpha*$ offset $\Delta \alpha*$ (mas)&${0.62}_{-0.31}^{+0.34}$&${0.27}_{-0.16}^{+0.17}$&0.62&0.38&$\mathcal{U}(-10^{6},10^{6})$\\
$\delta$ offset $\Delta \delta$ (mas)&${-0.99}_{-0.19}^{+0.18}$&${1.16}_{-0.19}^{+0.20}$&-0.96&1.19&$\mathcal{U}(-10^{6},10^{6})$\\
$\mu_{\alpha*}$ offset $\Delta \mu_{\alpha*}$ (mas\,yr$^{-1}$)&${-0.505}_{-0.056}^{+0.057}$&${-0.538}_{-0.062}^{+0.064}$&-0.484&-0.546&$\mathcal{U}(-10^{6},10^{6})$\\
$\mu_\delta$ offset $\Delta \mu_\delta$ (mas\,yr$^{-1}$)&${-0.107}_{-0.034}^{+0.037}$&${0.056}_{-0.035}^{+0.037}$&-0.094&0.055&$\mathcal{U}(-10^{6},10^{6})$\\
$\varpi$ offset $\Delta \varpi$ (mas)&\multicolumn{2}{c}{${-0.13}_{-0.11}^{+0.11}$}&-0.14&-0.17&$\mathcal{U}(-10^{6},10^{6})$\\
\hline
\textbf{Instrumental parameters}\\
RV offset for MIKE $\gamma_{\rm MIKE}$ (m\,s$^{-1}$)&\multicolumn{2}{c}{${-13.0}_{-2.9}^{+3.5}$}&-12.8&-14.6&$\mathcal{U}(-10^{6},10^{6})$\\
RV offset for PFS $\gamma_{\rm PFS}$ (m\,s$^{-1}$)&\multicolumn{2}{c}{${3.39}_{-0.58}^{+0.67}$}&3.24&3.67&$\mathcal{U}(-10^{6},10^{6})$\\
RV jitter for MIKE $J_{\rm MIKE}$ (m\,s$^{-1}$)&\multicolumn{2}{c}{${8.8}_{-1.8}^{+2.3}$}&8.0&7.5&$\mathcal{U}(0,10^{6})$\\
RV jitter for PFS $J_{\rm PFS}$ (m\,s$^{-1}$)&\multicolumn{2}{c}{${2.86}_{-0.48}^{+0.59}$}&2.59&2.32&$\mathcal{U}(0,10^{6})$\\
Jitter for hipparcos $J_{\rm hip}$ (mas)&\multicolumn{2}{c}{${3.7}_{-1.2}^{+1.1}$}&2.6&3.2&$\mathcal{U}(0,10^{6})$\\
Error inflation factor $S_{\rm gaia}$&\multicolumn{2}{c}{${1.124}_{-0.072}^{+0.084}$}&1.113&1.172&$\mathcal{N}(1,0.1)$\\
\hline
Mutual inclination $\psi_{bc}$ ($^\circ$)&${5.0}_{-2.2}^{+3.7}$&${162.1}_{-4.7}^{+7.1}$&3.2&168.5&--\\
\hline
\multicolumn{6}{l}{$^*$ $\chi^2$ statistics: $\chi_{\rm RV}^2=48$ ($\chi^2_{\rm reduce}=1.17$), $\chi_{\rm hip}^2=102$, $\chi_{\rm gaia}^2=16$. }\\
\multicolumn{6}{l}{$\rm ^a$ The maximum a posteriori (MAP) value. The global MAP value corresponds to prograde scenario, i.e., $i_c<90^\circ$.}.\\
\multicolumn{6}{l}{$\rm ^b$ Log-$\mathcal{U}(a, b)$ is the logarithmic uniform distribution between $a$ and $b$, Cos$i$-$\mathcal{U}(a, b)$ is the cosine uniform distribution of $i$}\\
\multicolumn{6}{l}{between $a$ and $b$, and $\mathcal{N}(a, b)$ is the Gaussian distribution with mean $a$ and stardare deviation $b$.}\\
\multicolumn{6}{l}{$\rm ^c$ The argument of periastron of the stellar reflex motion, differing by $\pi$ with planetary orbit, i.e., $\omega_{\rm p}=\omega+\pi$.}.\\
\end{tabular*}
\end{table*}

We adopt uniform priors for most fitting parameters. Specifically, the stellar mass is assigned a Gaussian prior with the value derived from our SED fit, and the inclination of HD 164604 b is also given a Gaussian prior based on Gaia two-body solution. With the total likelihood, $\mathcal{L}=\mathcal{L}_{\rm RV}\cdot\mathcal{L}_{\rm hip}\cdot\mathcal{L}_{\rm gaia}$, we finally derive the orbital solution by sampling the posterior via the parallel-tempering Markov Chain Monte Carlo (MCMC) sampler \texttt{ptemcee} \citep{Vousden2016}. We employ 30 temperatures, 100 walkers, and 80,000 steps per chain to generate posterior distributions for all the fitting parameters, with the first 40,000 steps being discarded as burn-in. We begin with an RV-only fit to obtain rough estimates of $P$, $K$, $e$, $\omega$ and $M_0$. These values will be set as initial conditions for subsequent joint fit, while the initial values of $i$, $\Omega$, $\Delta \alpha*$, $\Delta \delta$, $\Delta \mu_{\alpha*}$ and $\Delta \mu_\delta$ are drawn from uniform priors to explore possible bimodal solutions.

It should be noted that the smearing effect is inherent to our approach and to any similar works that relies on Hipparcos–Gaia absolute astrometry to constrain companion orbits. 
Because individual Gaia epoch measurements are unavailable, all positional information is compressed into the five astrometric parameters, so smearing becomes significant for companions whose orbital periods are comparable or shorter than Gaia observational baseline($\sim1000$\,days).
Therefore, previous studies typically restricted their method to long-period systems or neglected the astrometric signal of shorter-period companions (e.g., \citealt{Li2021}). 
The latter only works when the signal is far below the Gaia detection limit; for systems with detectable signals, the results may be biased.
With multiple Gaia releases, short-scale astrometric variations might now be captured by the discrepancy between GDR2 and GDR3, thus enabling constraints on short-period companions.

In our analysis, we did not apply a priori corrections for reference frame or other systematics between Hipparcos, GDR2 and GDR3 (e.g.,\citealt{Lindegren2020A&A, Brandt2021}). Instead, we absorbed them into the model via jitter terms ($J_{\rm hip}$, $S_{\rm gaia}$) and offsets ($\Delta \alpha*$, $\Delta \delta$, $\Delta \varpi$, $\Delta \mu_{\alpha*}$ and $\Delta \mu_\delta$). The latter not only determine the system’s barycenter but also absorb systematics in the Hipparcos and Gaia data. Sensitivity tests by \citet{Feng2021MNRAS} (see their Section 5) show that such corrections have negligible influence on the derived orbits. To confirm this, we repeated the fit after implementing the frame rotation calibration recommended by \citet{Feng2024ApJS}\footnote{A Python script is available in \url{https://github.com/ruiyicheng/Download_HIP_Gaia_GOST}}. No evident differences in the orbital solutions were found, so we omit this correction. Our model also implicitly treats GDR2 and GDR3 as independent. However, because $60–70\%$ of the GDR3 measurements strongly overlap with those in GDR2, this could bias our results. We quantify the effect by explicitly modelling the correlation between GDR2 and GDR3 following the derivation of \citet{Thompson2025AJ}; again, we find no significant influence on the orbital solutions. A detailed discussion is presented in Appendix~\ref{APPE_C}.

\subsection{Orbital solutions}
\begin{figure*}[ht!]
    \centering
    \includegraphics[width=0.8\textwidth]{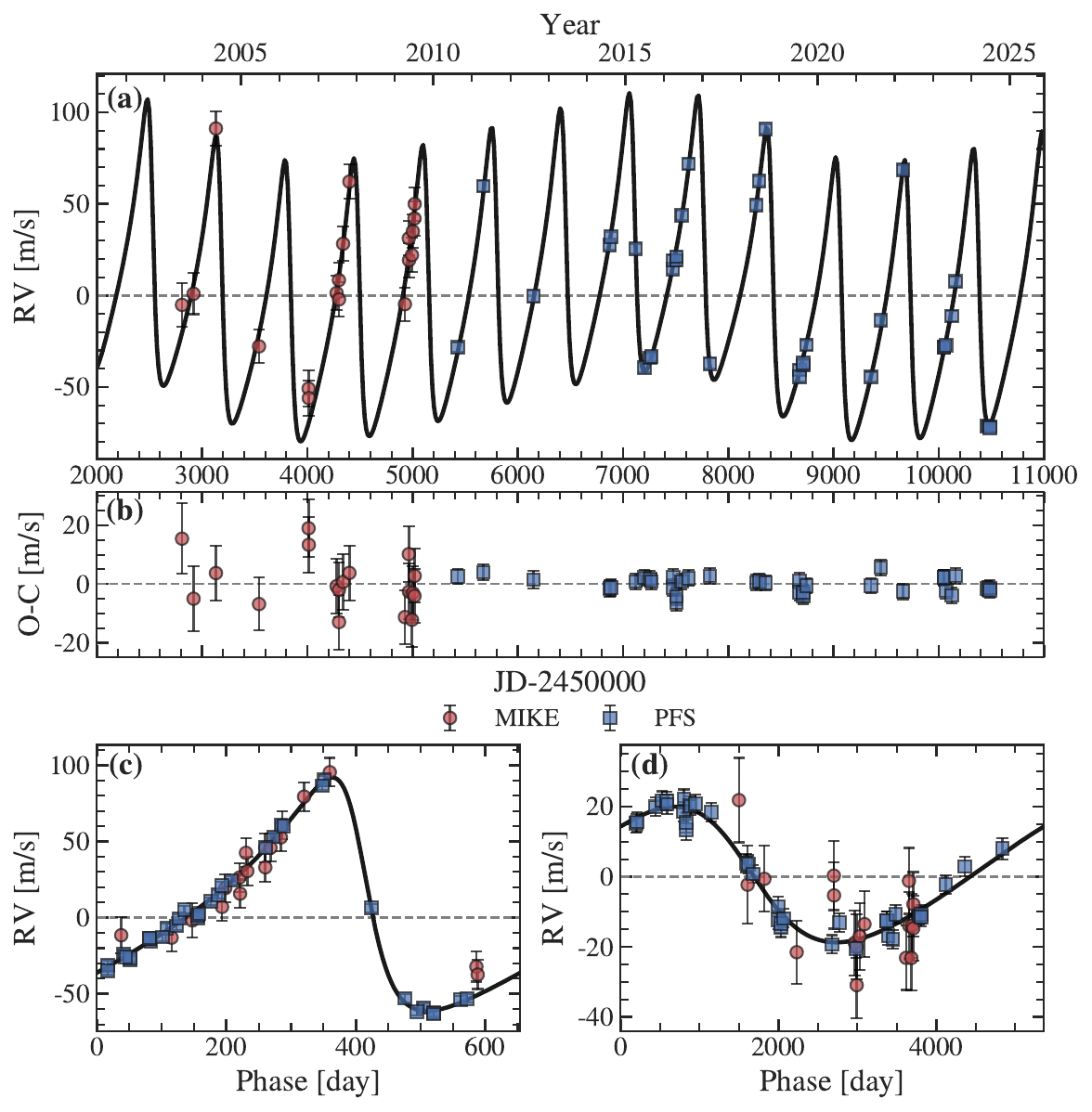}\caption{Best-fit orbit to HD\,164604 RVs from our joint fitting. \textbf{Panel (a)} shows the best-fit two-planet Keplerian orbit (thick black line) to the RV measurements from MIKE and PFS. \textbf{Panel (b)} displays the associated residuals. \textbf{Panel (c)} presents the phase-folded orbit of the inner planet HD\,164604\,b, with the signal of the outer planet HD\,164604\,c being subtracted. Similarly, \textbf{Panel (d)} shows the phase-folded orbit of the outer planet HD\,164604\,c, after correcting for the signal of HD\,164604\,b. 
\label{fig:rv_fit}}
\end{figure*}

\begin{figure*}[ht!]
    \centering
    \includegraphics[width=0.99\textwidth]{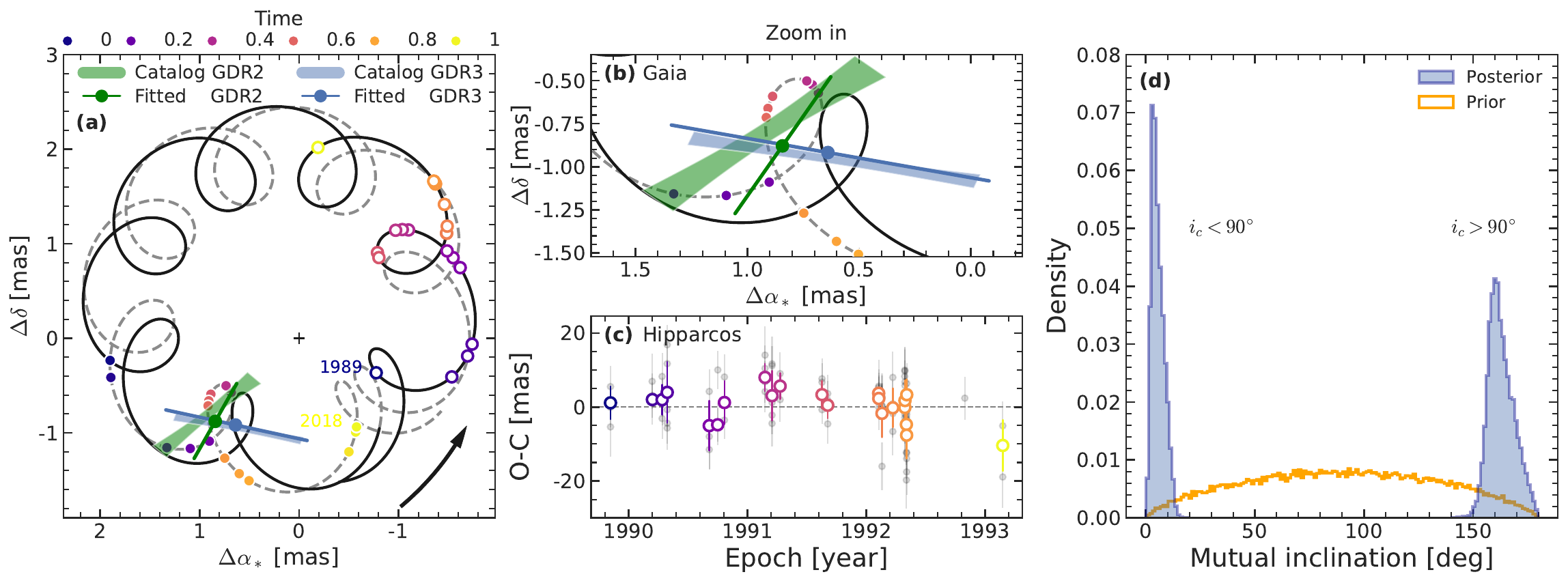}\caption{Best-fit orbit to the Hipparcos-Gaia astrometry of HD\,164604 from our joint fitting. \textbf{Panel (a)} plots the best-fit astrometric orbit of the star in the sky-projected plane. The black thick line and the gray dashed line denote the first and second periodic cycles of HD\,164604\,c, respectively. The central plus symbol marks the system's barycenter. Gaia epoch data simulated with GOST are indicated by colored solid dots, while the positions corresponding to the Hipparcos sampling epoch are denoted by hollow dots. The brightness of these points gradually increases with observation time (the temporal baseline of each satellite is set to 1). The arrow indicates the direction of the star's orbital motion. \textbf{Panel (b)} is an enlargement of the region of the fitting to Gaia astrometry, depicting the best fit to Gaia GOST data and the comparison between the best-fit and catalog astrometry (positions and proper motions) at GDR2 and GDR3 reference epochs. The shaded regions represent the uncertainty of catalog positions and proper motions after removing the motion of the system's barycenter. The two segments and their center dots (green and blue) represent the best-fit proper motion and position offsets induced by the planets for GDR2 and GDR3, respectively. \textbf{Panel (c)} shows the residual (O-C) of the Hipparcos 1D abscissa. The gray dots represent the unbinned data points, while the colored dots are the binned points. \textbf{Panel (d)} displays the prior and the posterior distributions of the mutual inclination between HD\,164604\,b and c. The bimodal distribution arises from the ambiguity of the orbital orientation of HD\,164604\,c.   
\label{fig:ast_fit}}
\end{figure*}

In this section, we present the orbital parameters of both planets derived from our joint RV–astrometry analysis. 
Table~\ref{Tab:result} lists the detailed parameters from MCMC chains, Figure~\ref{fig:rv_fit} shows the best fit to RVs, and Figure~\ref{fig:ast_fit} depicts the best fit to Hipparcos-Gaia astrometry. 
We adopt the posterior median as the best-fit value and quote the 16th–84th percentile range ($1\sigma$) as the uncertainty.
The astrometric acceleration of HD\,164604 and the posterior distributions of primary fitted parameters are presented in Appendix~\ref{APPE_B}. We find that the posteriors of $i_b$, $i_c$, $\Omega_c$, $\Delta \alpha*$, $\Delta \delta$, $\Delta \mu_{\alpha*}$ and $\Delta \mu_\delta$ exhibit evident bimodality, indicating that the current astrometry is insufficient to simultaneously determine the true orbital orientation of two non-transiting planets. We therefore separately report their two alternative solutions based on $i_c$. This ambiguity will be resolved with the release of future Gaia epoch astrometric data. The Bayesian evidence differences of the two-planet astrometric model over the one-planet and zero-planet model are $\Delta \ln Z=28$ and $11$, respectively, underscoring the value of adding astrometric constraints.

\subsubsection{Revise the orbit of HD\,164604\,b}
\begin{figure}[ht!]
    \centering
    \includegraphics[width=0.4\textwidth]{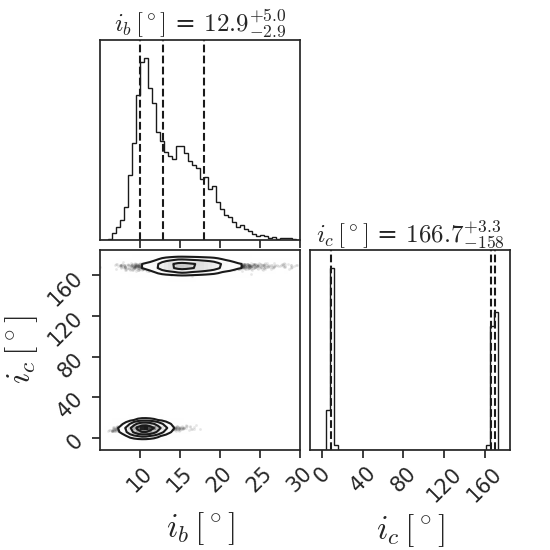}\caption{Correlation between the inclinations of the inner planet and the outer planet. 
\label{fig:inc1_inc2}}
\end{figure}

HD\,164604\,b was first reported by \citet{Arriagada2010ApJ} based on 18 MIKE RVs. They found the planet has a semi-major axis of $1.3\pm0.05\,\rm au$, an eccentric of $0.24\pm0.14$ and a minimum mass of $2.7\pm1.3\,M_{\rm Jup}$. 
\citet{Feng2019ApJS..242...25F} refined the orbit with 19 PFS RVs, obtaining a slightly higher eccentricity of $0.35\pm0.10$. More recently, Gaia-only astrometry further determined the mass of the planet to be $14.3\pm5.5\,M_{\rm Jup}$ and the inclination to be $29^\circ\pm19^\circ$, but left the eccentricity poorly constrained ($0.61\pm0.34$) \citep{Gaia_two_body_2023}. All of these studies adopted a single-planet model, which might be responsible for the large eccentricity uncertainty or the underestimate. Our joint fit updates the orbital parameters of HD\,164604\,b to $a_b={1.362}_{-0.012}^{+0.012}$\,au, $e_b={0.479}_{-0.021}^{+0.027}$, and $i_b={10.6}_{-1.3}^{+1.4}$$^\circ$ (or ${16.0}_{-2.8}^{+3.5}$$^\circ$). Our eccentricity is slightly higher than that of \citet{Arriagada2010ApJ}, but agrees with two other works within $1\sigma$. The inclination $i_b$ is found to be slightly bimodal and correlated with $i_c$ (see Figure~\ref{fig:inc1_inc2}), leading to two alternative mass estimates for the planet: ${13.2}_{-1.5}^{+1.8}\,M_{\rm Jup}$ or ${8.8}_{-1.5}^{+1.9}\,M_{\rm Jup}$. The two inclinations agree within $1.4\sigma$, indicating the orbital orientation of HD\,164604\,b is largely determined.
Both inclinations and masses align well with the Gaia-only values within $1\sigma$.

\begin{figure}[ht!]
    \centering
    \includegraphics[width=0.35\textwidth]{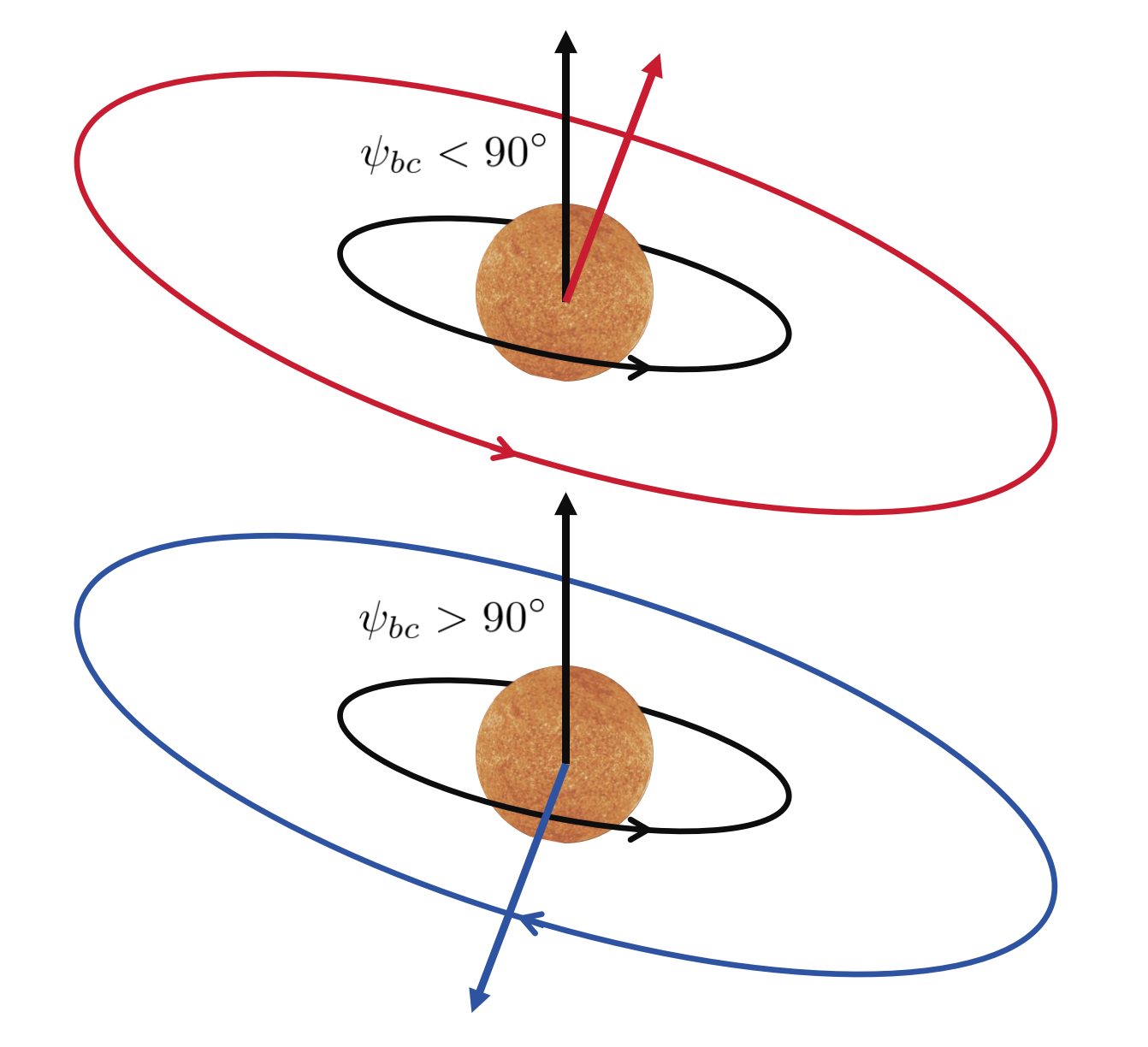}\caption{Orbital architectures of HD\,164604 system for prograde ($\psi_{bc}<90^\circ$) and retrograde ($\psi_{bc}>90^\circ$) scenarios. The stellar obliquity is unknown.
\label{fig:orb_archi}}
\end{figure}

\subsubsection{New discovery of HD\,164604\,c}
HD\,164604\,c is a newly discovered planet initially identified from decade-long PFS RV measurements that span nearly a complete orbital phase.
Our combined analysis reveals that the planet is a long-period super-Jupiter with a semi-major axis of ${5.556}_{-0.10}^{+0.093}$\,au, and an eccentricity of ${0.196}_{-0.078}^{+0.078}$ and a mass of ${9.5}_{-1.2}^{+1.2}\,M_{\rm Jup}$ (or ${7.6}_{-1.0}^{+1.0}\,M_{\rm Jup}$). We find two possible inclinations: ${8.6}_{-1.0}^{+1.3}$$^\circ$ and ${169.3}_{-1.7}^{+1.3}$$^\circ$. This degeneracy, which has been widely reported in previous studies that used Hipparcos–Gaia absolute astrometry to constrain long-period companions, becomes even more pronounced in multi-planet systems \citep{Feng2025MNRAS}. Taking 14\,Her as an example, this system consists of two long-period giant planets: 14\,Her\,b ($a_b=2.8$\,au and $m_b=8.9\,M_{\rm Jup}$) and c ($a_b=20.0$\,au and $m_b=7.9\,M_{\rm Jup}$). Through a joint analysis of RVs, Hipparcos-Gaia absolute astrometry and relative astrometry of JWST, \citet{Bardalez2025ApJ} recently found two possible inclinations for the inner planet (${146.7}_{-9.7}^{+4.6}$$^\circ$ and ${32.7}_{-4.0}^{+7.4}$$^\circ$). By incorporating GDR2, \citet{Xiao2025RNAAS} further constrained the orbital orientation of 14\,Her\,b, yielding a definitive inclination of $i_{b}={147.3_{-2.7}^{+2.2}}^{\circ}$. Therefore, additional astrometric data are essential to break this degeneracy.

\subsubsection{Mutual inclination between HD\,164604\,b and c}
In principle, the mutual inclination between two planets is determined by their individual inclinations and longitudes of the ascending nodes. The former denotes the angle between a planet's orbital angular-momentum vector and the observer's line of sight ($+\hat{z}$), and the latter represents the angle between the line of the ascending node and the reference direction ($+\hat{y}$). The true mutual inclination between planets can be expressed as
\begin{equation}
\cos \psi_{bc}=\cos i_b \cos i_c + \sin i_b \sin i_c \cos(\Omega_b-\Omega_c).
\end{equation}
As shown in panel (d) of Figure~\ref{fig:ast_fit}, the mutual inclination $\psi_{bc}$ exhibits a bimodal distribution, peaking at $3.2^\circ$ and $168.5^\circ$, corresponding to prograde ($\psi_{bc}<90^\circ$) and retrograde ($\psi_{bc}>90^\circ$) orbital configurations, respectively. Because the orbital orientation of the inner planet HD\,164604\,b is already well constrained, this bimodality reflects the remaining degeneracy in the orbit of HD\,164604\,c. Consequently, we derive two equally viable solutions for $\psi_{bc}$: either ${{5.0}_{-2.2}^{+3.7}}^\circ$ ($i_c<90^\circ$) or ${{162.1}_{-4.7}^{+7.1}}^\circ$ ($i_c>90^\circ$). The former is consistent with a nearly coplanar scenario ($1.7\sigma$ from $0^\circ$), whereas the latter implies a significant misalignment between the two orbits ($42\sigma$ from $0^\circ$ or $4.7\sigma$ from $180^\circ$). The orbital configurations are illustrated in Figure~\ref{fig:orb_archi}.

\section{Discussion} 
\label{sec:dis}
\subsection{Orbital stability}
\begin{figure*}[ht!]
    \centering
    \includegraphics[width=0.99\textwidth]{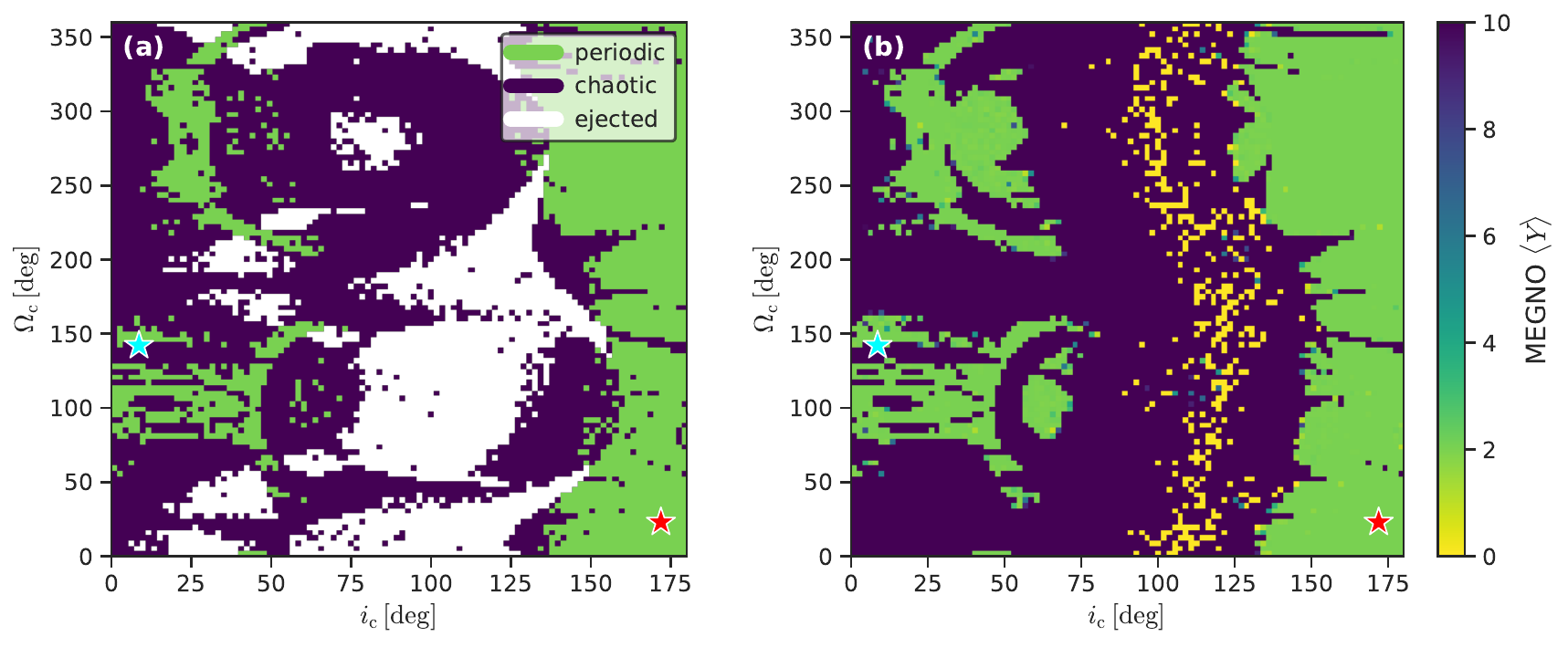}\caption{\texttt{MEGNO} simulation results of our Set $\mathrm{I}$ across different inclinations and ascending nodes of the outer planet HD\,164604\,c. \textbf{Panel (a)} is the stability map of three kinds of orbital evolutions. Green area indicates periodic evolution according to the \texttt{MEGNO} criterion, white area indicates one planet (most cases are HD\,164604\,c) was ejected, and dark blue area indicates chaotic evolution (not necessarily unstable).
    \textbf{Panel (b)} displays the \texttt{MEGNO} map of the same parameter space. Two possible values of $i_c$ and $\Omega_c$ are marked with cyan and red stars, respectively. The dynamically viable orbits are mainly located in the high $i_c$ region, corresponding to the  retrograde configuration.
\label{fig:megno_set1}}
\end{figure*}

To evaluate the dynamical stability of HD\,164604 system, we perform two sets of N-body simulations using the \texttt{MEGNO} (Mean Exponential Growth of Nearby Orbits, $\langle Y \rangle$) chaos indicator \citep{Cincotta2000A&AS} implemented in the package \texttt{REBOUND} \citep{Rein2012A&A}. In multi-planetary systems, the \texttt{MEGNO} indicator effectively discriminates between quasi-periodic and chaotic orbital evolution: a chaotic state strongly compromises a planet's long-term orbital stability \citep{Hinse2010MNRAS}. For stable periodic evolution, $\langle Y \rangle$ is close to 2, while for chaotic evolution, $\langle Y \rangle$ deviates from 2 \citep{Cincotta2000A&AS, Gozdziewski2001A&A, Hinse2010MNRAS}.
In Set $\mathrm{I}$, we fix most parameters to the global MAP values obtained from our joint fitting and uniformly sample the inclination $i_c$ across (0,$\pi$) and the longitude $\Omega_c$ across (0,$2\pi$) ($100\times100$ grids). This set explores the full range of possible orbital configurations for HD\,164604\,c, assuming its orbital orientation is unknown. In Set $\mathrm{II}$, we initialize 10,000 orbits randomly drawn from the MCMC chains to assess their long-term orbital stability. We choose \texttt{WHFAST} \citep{Rein2015MNRAS}, a symplectic Wisdom–Holman integrator to integrate 10 million years for each condition with a time step of 0.045 yr ($\sim16$ days).
We classify motions as periodic when $|\langle Y \rangle-2|<10^{-2}$, as chaotic (both planets remain physically bound) when $|\langle Y \rangle-2|>10^{-2}$, and as ejected when $a_b<0$ or $a_c<0$.


Figure~\ref{fig:megno_set1} presents the results of Set $\mathrm{I}$. Panel (a) illustrates the orbital evolution map of systems in the $i_c-\Omega_c$ space, with colors indicating periodic (green), chaotic (dark blue) and ejected evolutions (white). Panel (b) is color-coded based on the specific \texttt{MEGNO} values. We find that for initial configurations with $i_c\lessapprox130^\circ$, the orbits are predominantly chaotic or ejected within the integration time, and the stable orbits are confined to relatively narrow regions. Most of the stable periodic orbits are located in the high-inclination region, indicating a retrograde orbital configuration between HD\,164604\,b and c.

\begin{table}
\centering
\caption{Statistics for stability analysis of Set $\mathrm{II}$}\label{Tab:stable}
\begin{tabular}{lcc}
\hline \hline
 Stage & Prograde&Retrograde\\
\hline
Periodic &728&4588\\
Chaotic &2168&734\\
Ejected &1684&98\\
\hline
Total&4580&5420\\
\hline
\end{tabular}
\end{table}

\begin{figure*}[ht!]
    \centering
    \includegraphics[width=0.95\textwidth]{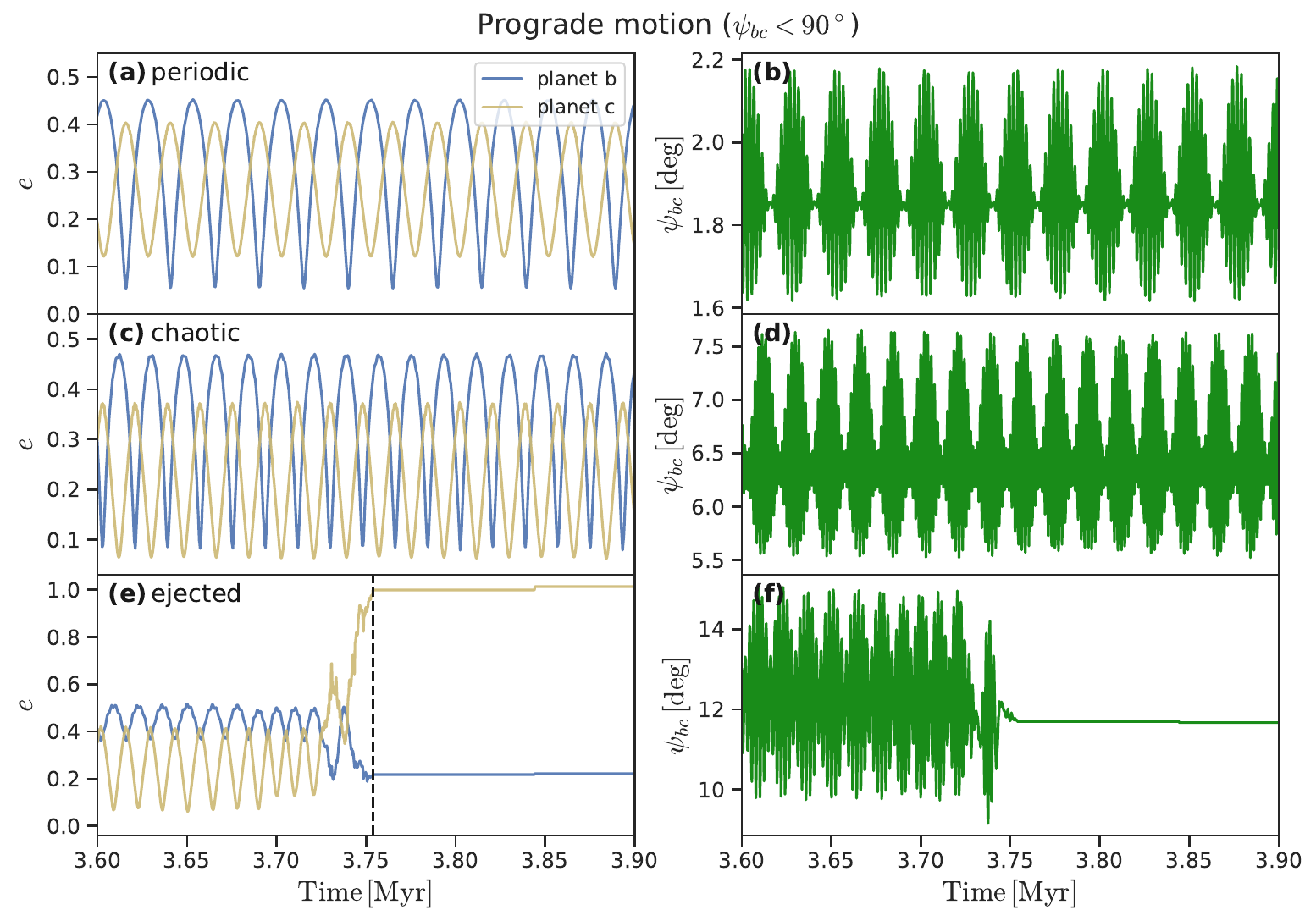}\caption{Example evolutions of HD\,164604\,b and c for prograde scenario ($\psi_{bc}<90^\circ$). \textbf{Panel (a) (c) (e)} respectively show the periodic, chaotic and ejected evolutions of eccentricity, and \textbf{Panel (b) (d) (f)} show their corresponding mutual inclinations. The ejection of the outer planet HD\,164604\,c occurs at $t \sim 3.75$ Myr, where $e_c > 1$ (vertical dashed line). The x-axis is truncated for visualization purposes.
\label{fig:prograde}}
\end{figure*}

\begin{figure*}[ht!]
    \centering
    \includegraphics[width=0.95\textwidth]{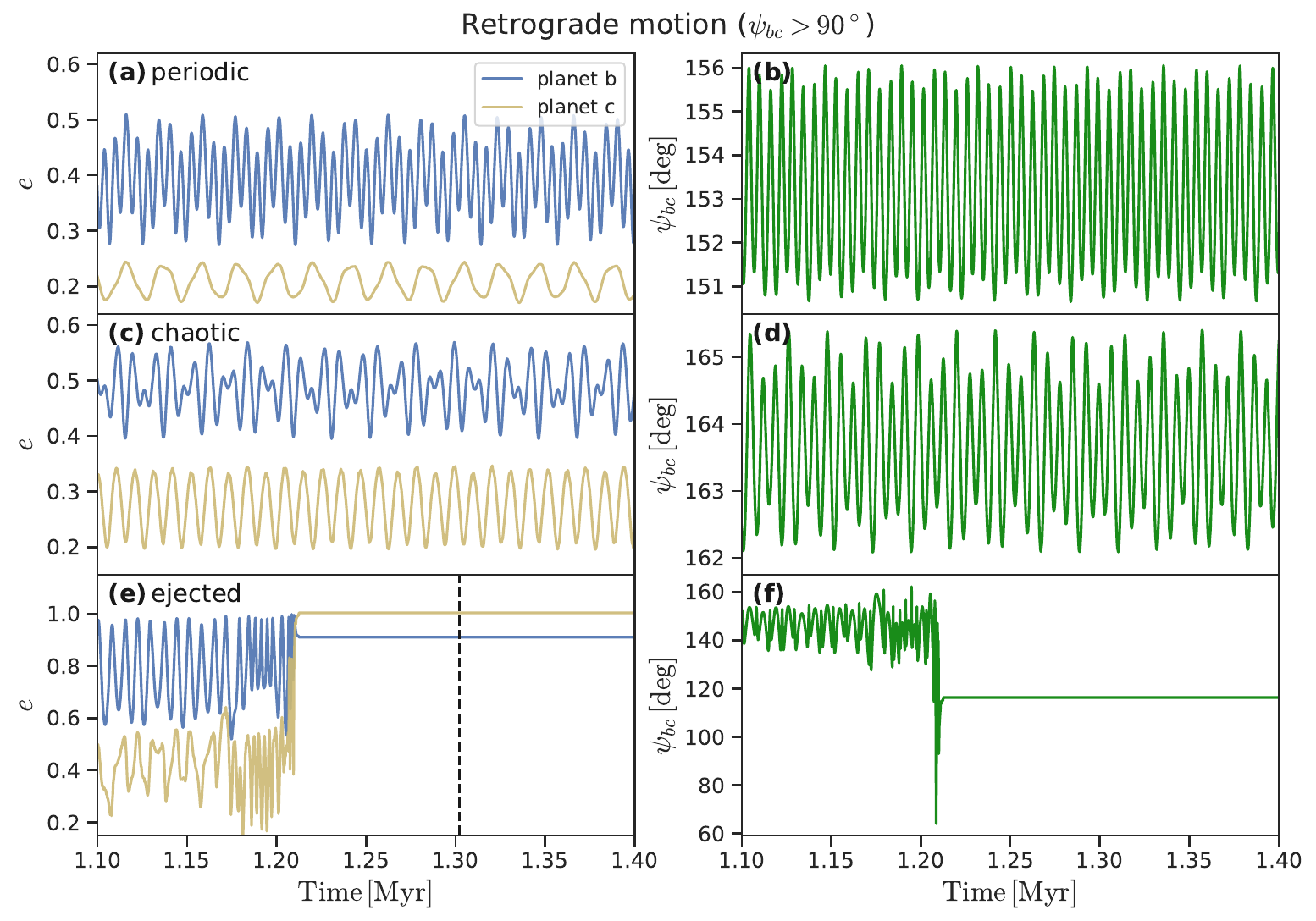}\caption{Example evolutions of HD\,164604\,b and c for retrograde scenario ($\psi_{bc}>90^\circ$). The meanings of each panel are the same as those in Figure~\ref{fig:prograde}.
\label{fig:retrograde}}
\end{figure*}

Table~\ref{Tab:stable} presents the summary of the simulation results of Set $\mathrm{II}$. For the prograde scenario, we find that only $15.9\%$ of orbits can maintain a stable periodic state over 10 million years, while $47.3\%$ of orbits exhibit chaotic behavior, and $36.8\%$ of orbits will result in the ejection of the outer planet. 
For the retrograde scenario, the majority of orbits ($84.6\%$) remain periodically stable throughout the integration time, while chaotic and ejected orbits account for only $13.5\%$ and $1.8\%$, respectively. 
Because retrograde orbits produce larger relative velocities and shorter encounter times than prograde orbits, the resulting reduction in cumulative angular-momentum exchange suppresses long-term chaotic diffusion, thereby extending the system’s stability to much longer timescales. Analogous cases include the HD\,47366\citep{Sato2016ApJ} and BD+20 2457\citep{Horner2014MNRAS} systems.
Statistically, the orbital architecture of the HD\,164604 system appears to favor a retrograde configuration, but the possibility of a prograde scenario can not fully exclude based on current observations. Two example evolutions for these two scenarios are presented in Figure~\ref{fig:prograde} and \ref{fig:retrograde}. 

\subsection{Formation scenario}
It is generally expected that planets should emerge on circular, coplanar orbits — much like those in our own Solar System — because strong gravitational coupling between the planets and their protoplanetary disk efficiently damps any initial eccentricities and inclinations. However, observations have uncovered numerous planetary systems whose orbits are moderately or extremely eccentric and strongly misaligned, implying a dynamically active formation history (e.g., \citealt{Lu2025ApJ}). Herein, we separately discuss the possible formation mechanisms of HD\,164604 planetary system according to its two possible orbital configurations.

For the prograde scenario, the measured mutual inclination ($\psi_{bc}=5^\circ$) between HD\,164604\,b and c is consistent with a coplanar configuration, suggesting that they did not experience a strong gravitational perturbation. Both planets might form inside the protoplanetary disk, and then reach their currently observed locations by the disk-driven migration. Their moderate eccentricities could be excited by several possible mechanisms, such as planet-planet scattering\citep{Ford2008}, secular chaos\citep{Wu2011ApJ}, and planet-disk interactions\citep{Goldreich2003ApJ}. Planet-planet scattering may occur in the early stage of the formation of tightly packed multiple giant planets. A third, lighter planet might be ejected by the gravitational instability (e.g., close encounters) and left the remaining pair with significant eccentricities but a mutual inclination still below $\sim5^\circ$. Secular chaos can effectively pump the eccentricity of the innermost planet in multi-planetary systems and is commonly used to explain the formation of hot Jupiters. However, in strictly coplanar systems, it might be hard to work without a distant and inclined companion. The last one, planet-disk interactions, is sensitive to the planet's initial eccentricity, gap openning of gaseous disk, and the disk properties.

For the retrograde scenario, a far more violent dynamical route is required to flip the planetary orbits to $\psi_{bc}\sim162^\circ$. The von-Zeipel–Lidov–Kozai (ZLK; \citealt{vonZeipel1910AN, Kozai1962, Lidov1962}) mechanism offers a promising pathway to produce large eccentricities and mutual inclinations, but it generally needs an initial misalignment of $39.2^\circ<\psi<140.8^\circ$. Assuming an unsolved perturber on a wider, moderately inclined orbit, high-order ZLK effect might simultaneously excite the eccentricities of both planets and their mutual inclination, even reversing the orientation of the inner orbit. Another promising channel is three-body scattering\citep{Ford2008} during the system's early evolution. In this paradigm, a comparably massive planet is either ejected or scattered into the host star, the angular momentum of the remaining two planets redistributed, and their mutual inclination might be reversed. However, because the stellar obliquity and the exact orbital configuration of the outer planet remain unknown, it is still too early for us to draw a definitive conclusion about their formation.

\subsection{Systems with mutual inclination measurement}
\begin{figure}[ht!]
    \centering
    \includegraphics[width=0.45\textwidth]{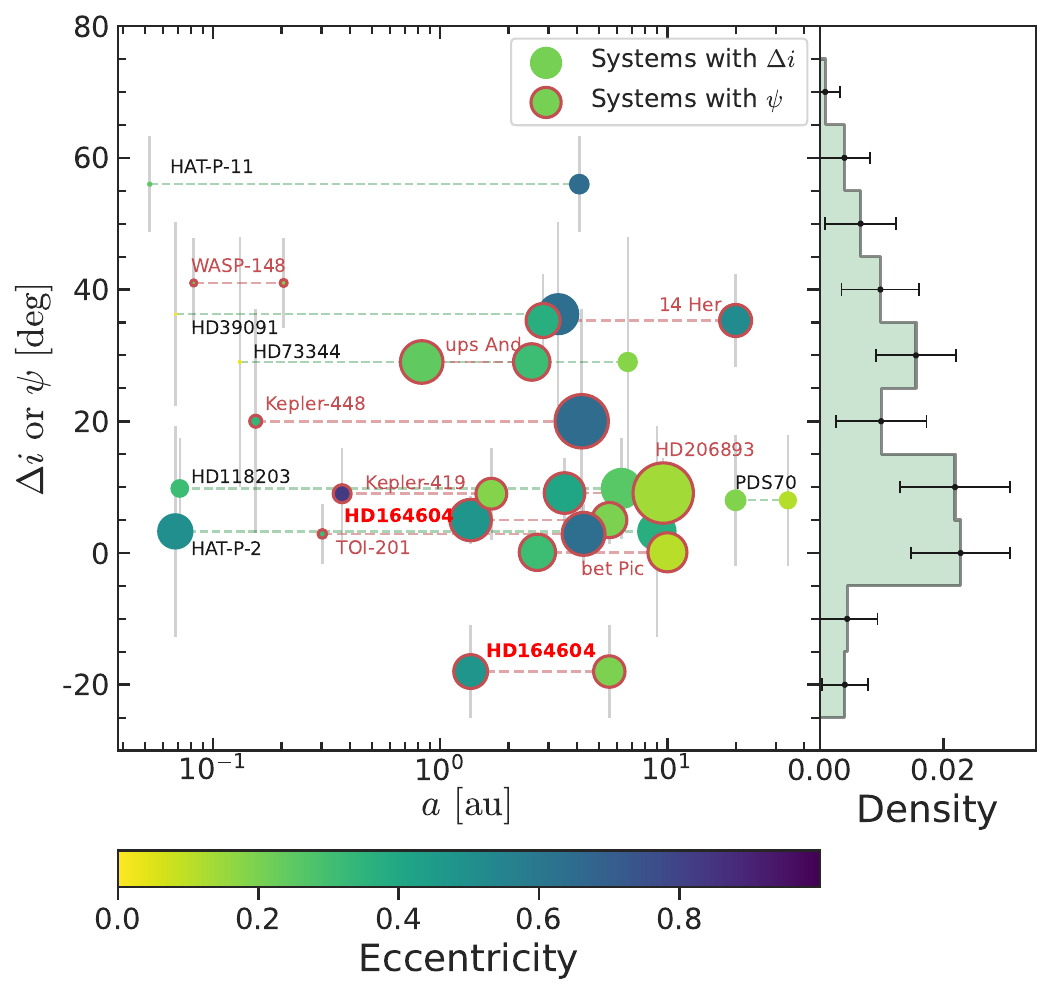}\caption{Systems with well constrained line-of-sight mutual inclination ($\Delta i$) or true mutual inclination ($\psi$). The color of each circle denotes eccentricity, and the size is proportional to the planet’s mass. Transiting compact systems are not included. Right panel is the 1D distribution after taking into account the observed uncertainties. 
    Data references and relevant methods are provided below. RV+Hip-Gaia astrometry: HAT-P-11 \citep{An2025AJ,Basilicata2024A&A}; HAT-P-2 \citep{de_Beurs2023AJ}; HD\,39091 \citep{Huang2018ApJ,Xuan2020MNRAS,Feng2022ApJS}; HD\,118203 \citep{Zhang2024AJ}; HD\,73344 \citep{Zhang2025AJ}. TTV: WASP-148 \citep{Almenara2022A&A}; Kepler-448 \citep{Masuda2017AJ}; Kepler-419 \citep{Dawson2014ApJ}; TOI-201 \citep{Maciejewski2025}. RV+Hip-Gaia+imaging: bet\,Pic \citep{Lagrange2019NatAs,Feng2022ApJS}; 14\,Her \citep{Bardalez2025ApJ,Xiao2025RNAAS}. RV+HST astrometry: ups\,And \citep{McArthur2010ApJ}. RV+imaging: HD\,206893 \citep{Hinkley2023A&A}. Imaging: PDS70 \citep{Wang2020AJ}.
\label{fig:samp}}
\end{figure}

Figure~\ref{fig:samp} depicts multi-planet systems for which the line-of-sight mutual inclination ($\Delta i$) or the true mutual inclination ($\psi$) is well constrained. Here $\Delta i$ is defined as $|i_{\rm in}-i_{\rm out}|$, which represents the lower limit of $\psi$. 
Because the longitude of the ascending node of the inner transiting planet is usually unknown, only the absolute difference between $i_{\rm in}$ and $i_{\rm out}$ can be measured directly (e.g., HD\,73344, \citealt{Zhang2024AJ}). 
In transiting systems that exhibit large transit-timing variations (TTVs), the true $\psi$ can be derived with TTV dynamical modeling (e.g., Kepler-448, \citealt{Masuda2017AJ}). 
For non-transiting systems, constraints form astrometry (e.g., HST, Hip-Gaia and direct imaging) are specially essential (e.g., ups\,And, \citealt{McArthur2010ApJ}, 14\,Her, \citealt{Bardalez2025ApJ}). 
The figure shows that roughly half of systems are consistent with nearly coplanar orbits ($<10^\circ$), especially those hosting two comparably massive giants, whereas significantly misaligned systems remain rare. Assuming both planets in HD\,164604 are on prograde orbit, we find the system exhibits a similar architecture as bet Pic system\citep{Lagrange2019NatAs}: both host very massive planets at relatively wide separations and share the same eccentricity pattern ($e_{\rm in}>e_{\rm out}$), implying their relatively mild formation and evolution histories. However, a confirmed retrograde configuration would imply a far more complex dynamical past of HD\,164604 system. Detailed dynamical modeling and a comprehensive population-level study that accounts for detection biases are therefore urgently required.

\subsection{Detectability with Gaia DR4}
Gaia DR4 (GDR4) is expected to release time-series astrometric data for more than a million stars in December 2026, based on 66 months of observations. To assess whether the two possible orbital configurations (prograde or retrograde) of the HD 164604 system can be differentiated using the longer-baseline data from GDR4, we simulate the expected abscissae  using the orbital elements derived from MAP values of both scenarios. We then calculate the Bayesian information criterion (BIC; \citealt{Schwarz1978}) difference, $\rm \Delta BIC$, for model comparison. The BIC is defined as $-2\,{\ln}L+k\,{\ln}\,n$, where ${\ln}L$ is log likelihood, $n$ is the number of data points, and $k$ is the number of free parameters.
We assume a prograde scenario (Model 0) for the system, i.e., ${\ln}L_0=0$. Therefore, for retrograde scenario (Model 1), we obtain ${\ln}L_1=-45$, yielding a BIC difference of $\rm \Delta BIC_{10}=-90$. This significant difference indicates that the epoch astrometry from GDR4 will be able to readily determine the true orbit of the HD\,164604 system. 

\begin{figure}[ht!]
    \centering
    \includegraphics[width=0.42\textwidth]{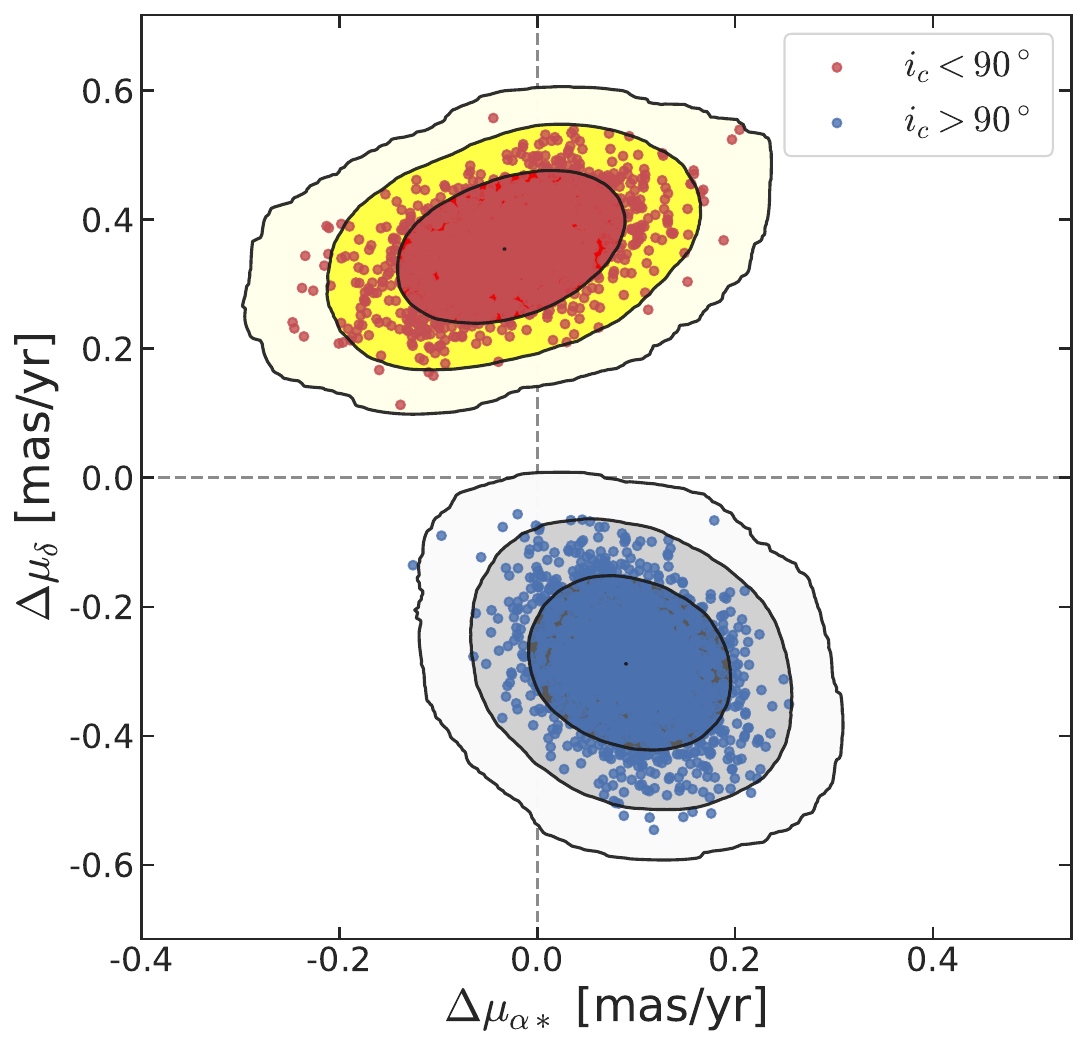}\caption{Predicted GDR4 proper motions relative to GDR3. Red and blue areas respectively represent the prograde ($i_{c}<90^\circ$) and retrograde ($i_{c}>90^\circ$) scenarios. The contour lines from the inside and out denote the $1\sigma$, $2\sigma$ and $3\sigma$ uncertainties.
\label{fig:DR4}}
\end{figure}

In addition, we predict the expected proper motions for these two scenarios using the standard five-parameter astrometric model. As shown in Figure~\ref{fig:DR4}, the difference in R.A. is small but in decl., it exceeds $3\sigma$. This suggests that we can use the catalog astrometry of Gaia DR4 to directly distinguish between these two scenarios.

\section{Conclusion} 
\label{sec:con}
We report the discovery of an additional long-period giant planet orbiting the K3.5 dwarf HD 164604 and determine the planet–planet mutual inclination for this non-transiting multi-planet system via a joint analysis of RV, Hipparcos, and Gaia DR2 and DR3 astrometry. This system thereby joins the small but growing population whose mutual inclinations are well characterized. We highlight the advantage of our approach for characterizing the orbital properties of cold Jupiters. Our main conclusions are summarized as follows.

\begin{enumerate}
\item We refine the orbit and dynamical mass of the inner giant planet HD\,164604\,b as: $a_b={1.362}_{-0.012}^{+0.012}$\,au, $e_b={0.479}_{-0.021}^{+0.027}$, $i_b={10.6}_{-1.3}^{+1.4}$$^\circ$ (or ${16.0}_{-2.8}^{+3.5}$$^\circ$) and $m_b={13.2}_{-1.5}^{+1.8}\,M_{\rm Jup}$ (or ${8.8}_{-1.5}^{+1.9}\,M_{\rm Jup}$). 

\item The outer long-period giant planet HD\,164604\,c has a semi-major axis of $a_c={5.556}_{-0.10}^{+0.093}$\,au, an eccentrity of $e_c={0.196}_{-0.078}^{+0.078}$, an inclination of $i_c={8.6}_{-1.0}^{+1.3}$$^\circ$ (or ${169.3}_{-1.7}^{+1.3}$$^\circ$) and a mass of $m_c={9.5}_{-1.2}^{+1.2}\,M_{\rm Jup}$ (or ${7.6}_{-1.0}^{+1.0}\,M_{\rm Jup}$). 

\item We constrain the true mutual inclination between these two giant planets to be $\psi_{bc}={5.0}_{-2.2}^{+3.7}$$^\circ$ or ${162.1}_{-4.7}^{+7.1}$$^\circ$ at $1\sigma$, respectively corresponding to prograde and retrograde orbital architectures. Our measurements demonstrate that the orbit of these two-planet system is either coplanar or anti-coplanar.

\item The stability analyses indicate that the prograde configuration of HD\,164604\,b and c primarily leads to either chaotic evolution or ejection of one of the planets, whereas retrograde orbits remain largely stable and quasi-periodic over the 10 Myr integration.

\item If coplanar ($\psi_{bc}\sim{5}$$^\circ$), the planets of the system likely formed within the disk and migrated inward or outward, with moderate eccentricities excited by weak scattering or disk torques. The retrograde case ($\psi_{bc}\sim{162}$$^\circ$) demands violent dynamical evolution such as the high-order ZLK cycle from an unseen wide companion or early three-body scattering that ejected a third planet. However, the lack of information on a distant perturber and stellar obliquity precludes a definitive conclusion regarding these channels.

\item The simulated observations of Gaia DR4 suggest that the ambiguity in the orbital orientation of HD\,164604\,c will be readily resolved, enabling a more detailed exploration of the dynamical history of the HD 164604 system.

\end{enumerate}

\begin{acknowledgments}
We thank the referee for his constructive suggestions, which have greatly improved this paper. This work is supported by the National Key R\&D Program of China, No. 2024YFA1611801, and No. 2024YFC2207700 by the National Natural Science Foundation of China (NSFC) under Grant No. 12473066, by the Shanghai Jiao Tong University 2030 Initiative, and by the China-Chile Joint Research Fund (CCJRF No. 2205). CCJRF is provided by Chinese Academy of Sciences South America Center for Astronomy (CASSACA) and established by National Astronomical Observatories, Chinese Academy of Sciences (NAOC) and Chilean Astronomy Society (SOCHIAS) to support China-Chile collaborations in astronomy.
\end{acknowledgments}





%
\facilities{MIKE, PFS, Hipparcos, Gaia}

\software{astropy \citep{2013A&A...558A..33A,2018AJ....156..123A,2022ApJ...935..167A},  
          scipy \citep{2020SciPy-NMeth}, matplotlib \citep{Hunter:2007}, corner \citep{corner} , rebound \citep{Rein2012A&A}.}


\appendix

\section{Additional tables}
\label{APPE_A}
PFS radial velocities are measured with an iodine absorption cell inserted in the converging beam before the spectrometer slit \citep{Butler_1996}. The superimposed iodine lines provide both a wavelength scale and a time-variable proxy for the instrument PSF, allowing the model to track changes in optics and illumination on all timescales. Each observed spectrum is modelled as the product of a high-resolution stellar spectrum and an iodine spectrum, convolved with a parametric PSF whose free parameters are the wavelength scale, PSF coefficients and stellar Doppler shift. PFS can achieve an RV precision of $0.5-1.0$ ${\rm m\,s^{-1}}$ for nearby, bright, and Chromospherically quiet stars. For HD\,164604, we have collected 35 precise RVs, $S$ index and $H$ index ($H_\alpha$) from April 2010 to June 2024 (Table~\ref{Tab:RV_164604}). Perspective acceleration is removed internally by the pipeline. RV and astrometric data are also available on \dataset[Zenodo]{https://doi.org/10.5281/zenodo.18172547}.

\begin{table*}[h]
\centering
\caption{PFS RVs for HD\,164604}\label{Tab:RV_164604}
\begin{tabular*}{\textwidth}{@{}@{\extracolsep{\fill}}cccccccccc@{}}
\hline \hline
BJD & RV [$\rm m\,s^{-1}$] & Error [$\rm m\,s^{-1}$] & $S$ index & $H_{\alpha}$ & BJD & RV [$\rm m\,s^{-1}$] & Error [$\rm m\,s^{-1}$] & $S$ index & $H_{\alpha}$\\
\hline
2455428.60205&-14.84&1.00&0.395&0.042&2458355.55738&104.33&0.97&0.410&0.041\\
2455671.84628&73.25&1.20&0.421&0.043&2458674.72378&-27.12&1.20&0.541&0.042\\
2456149.62309&13.14&1.84&0.347&0.042&2458674.73179&-30.98&1.24&0.484&0.042\\
2456871.61061&41.03&1.23&0.365&0.043&2458709.61387&-24.58&0.95&0.420&0.042\\
2456883.61869&45.56&1.27&0.344&0.042&2458709.62259&-23.32&1.02&0.435&0.042\\
2457119.84066&38.97&1.15&0.517&0.043&2458741.52643&-13.51&1.01&0.381&0.043\\
2457200.72942&-25.86&1.04&0.476&0.043&2459356.80572&-31.03&0.98&0.444&0.043\\
2457257.59100&-20.59&0.99&0.366&0.042&2459447.61183&0.00&0.97&0.456&0.042\\
2457267.53811&-19.75&1.30&0.359&0.043&2459660.85626&82.05&1.02&0.402&0.044\\
2457472.84177&27.91&1.22&0.385&0.042&2460047.87249&-14.71&1.06&0.377&0.044\\
2457476.91264&32.78&1.15&0.318&0.043&2460047.88007&-14.67&1.05&0.396&0.045\\
2457505.82699&32.75&1.70&0.535&0.044&2460067.82857&-13.60&1.15&0.361&0.044\\
2457505.83670&34.78&1.51&0.545&0.044&2460124.69241&2.32&1.04&0.399&0.044\\
2457557.75918&57.31&1.15&0.464&0.041&2460159.65050&21.14&1.04&0.409&0.044\\
2457622.57560&85.31&1.12&0.343&0.042&2460460.74339&-57.81&1.00&0.392&0.044\\
2457825.90633&-23.71&1.22&0.498&0.042&2460486.73344&-58.03&1.17&0.440&0.044\\
2458264.81709&62.91&0.91&0.375&0.042&2460486.74425&-58.82&1.04&0.421&0.044\\
2458292.71329&76.05&0.86&0.398&0.042&&&&&\\
\hline
\end{tabular*}
\end{table*}

\section{Additional figure} \label{APPE_B}
\begin{figure*}[ht!]
    \centering
    \includegraphics[width=0.6\textwidth]{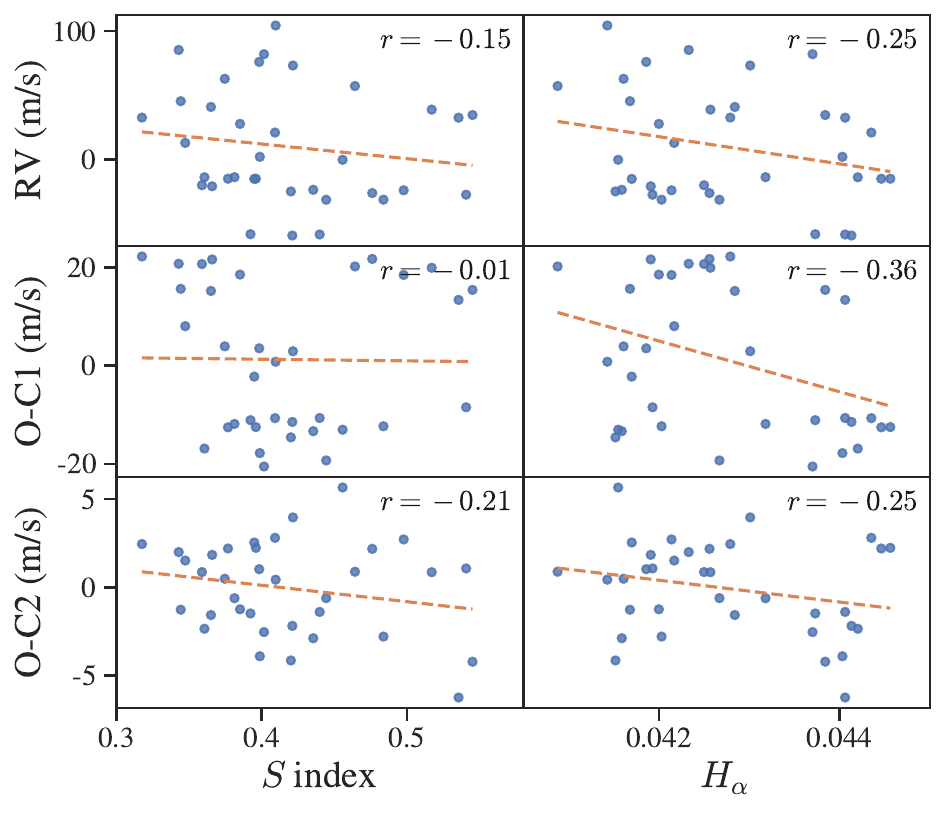}\caption{Correlation of the stellar-activity indices with RVs before and after removing the signal(s) of the inner planet (O-C1) and then both planets (O-C2). The Pearson's correlation coefficient $r$ is shown in each panel.  
\label{fig:correlation}}
\end{figure*}

\begin{figure*}[ht!]
    \centering
    \includegraphics[width=0.6\textwidth]{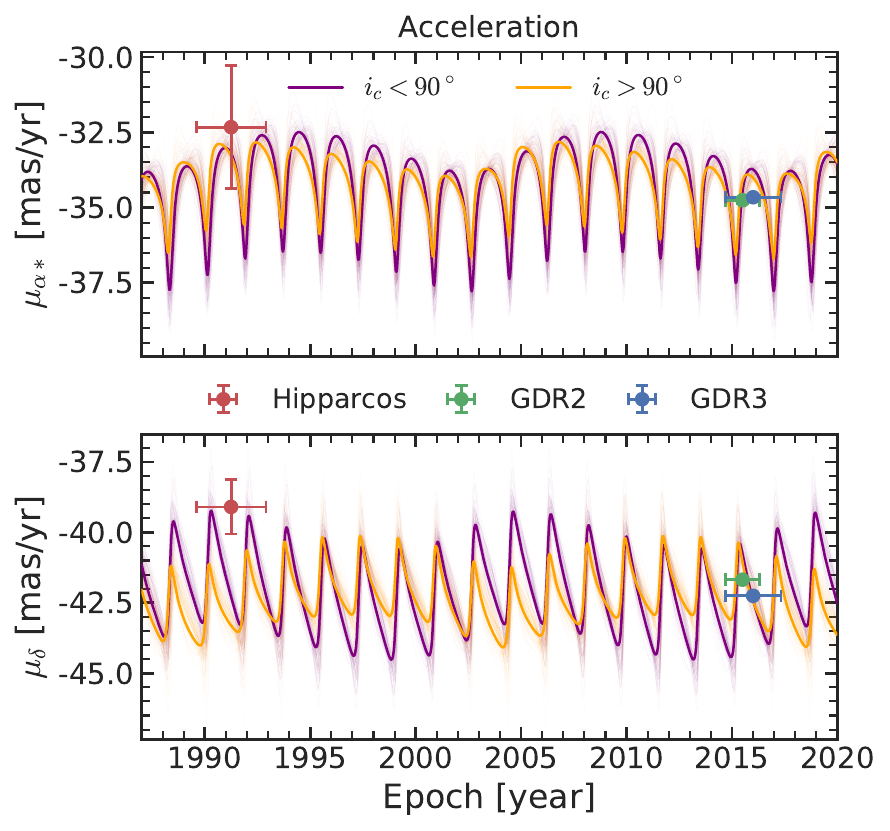}\caption{Astrometric acceleration of HD\,164604 in right ascension and declination. The black lines represent the best-fit orbit, and the colored lines indicate the possible orbital solution randomly drawn from the MCMC chain. The horizontal error bar of each data point denotes the observational baseline.
\label{fig:acceleration}}
\end{figure*}

\begin{figure*}[ht!]
    \centering
    \includegraphics[width=0.99\textwidth]{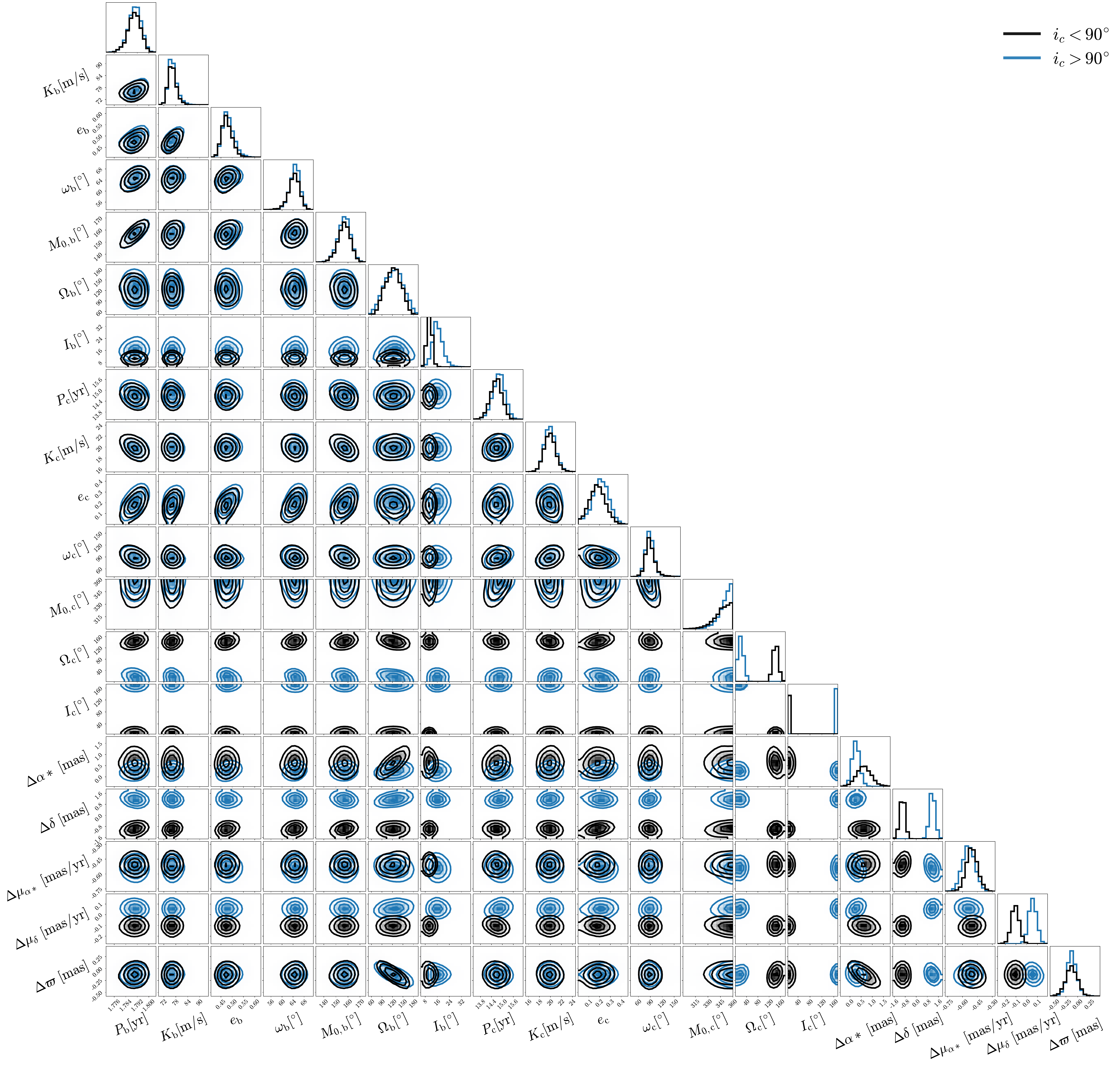}\caption{Posterior distributions of the selected orbital parameters for HD\,164604 system. 
\label{fig:corner}}
\end{figure*}

Figure~\ref{fig:correlation} shows the correlation of the stellar-activity indices with PFS RVs and residuals. Figure~\ref{fig:acceleration} shows the the fitting to astrometric acceleration of HD\,164604 in right ascension and declination, and Figure~\ref{fig:corner} shows the posterior distributions of the primary fitted parameters.

\section{Correlation between GDR2 and GDR3}
\label{APPE_C}
\begin{figure*}[ht!]
    \centering
    \includegraphics[width=0.99\textwidth]{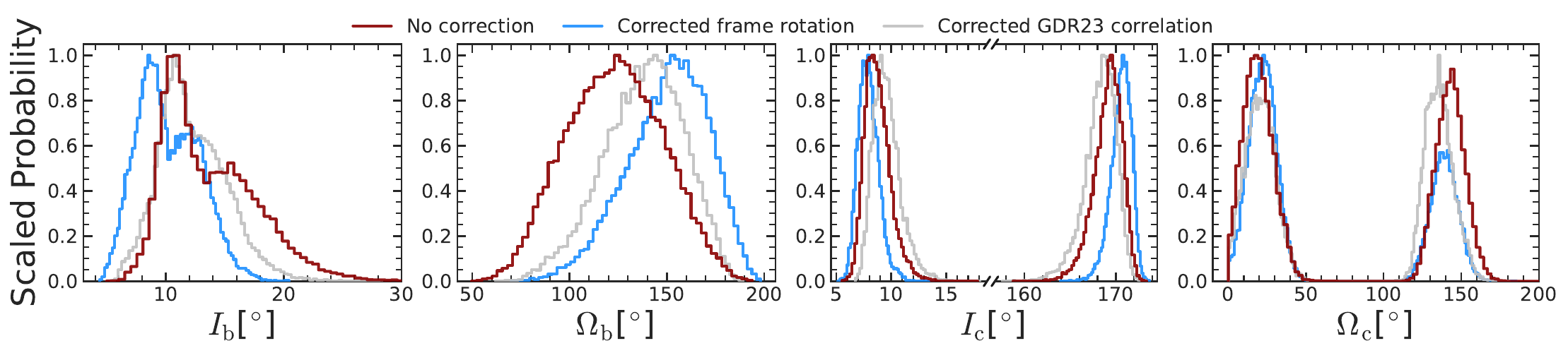}\caption{Comparison of orbital parameters between solutions with and without corrections to the reference frame and GDR23 correlation. Only inclination ($I$) and longitude of the ascending node ($\Omega$) are shown, as these two elements are primarily constrained by astrometric data. We find that the orbital solutions are not sensitive to these corrections. 
\label{fig:correction}}
\end{figure*}

When accounting for the correlation between GDR2 and GDR3, the likelihood becomes
\begin{equation}
  \mathcal{L}_{\rm gaia}=
  \frac{1}{\sqrt{(2\pi)^{10}|\Sigma(S^2)|}}\times
  {\rm exp}\left(-\frac{1}{2}(\Delta\hat{\vec{\iota}}-\Delta\vec{\iota})^{T}[\Sigma(S^2)]^{-1}(\Delta\hat{\vec{\iota}}-\Delta\vec{\iota})    \right)~.
\end{equation}
$\Delta\vec{\iota}$ is now a ten-parameter astrometric vector, and $\Sigma$ is the full covariance matrix ($10\times10$) between the two catalogs, written as
\begin{equation}
\Sigma=\begin{bmatrix}
\Sigma_{\rm DR2} & \rho\,\sqrt{\Sigma_{\rm DR2}}\,\sqrt{\Sigma_{\rm DR3}}\, \\
\rho\,\sqrt{\Sigma_{\rm DR2}}\,\sqrt{\Sigma_{\rm DR3}} & \Sigma_{\rm DR3}
\end{bmatrix},
\end{equation}
where $\rho$ is the corelation. \citet{Thompson2025AJ} derived $\rho=\sqrt{N_{\rm DR2}/N_{\rm DR3}}$ for the DR2-DR3 proper motion. Following their derivation, it is easy to prove that this expression still holds for position. For simplicity, we adopt it for parallax as well. According to GOST simulation, the number of GDR2 and GDR3 transits (after excluding satellite dead-time) for HD\,164604 is $N_{\rm DR2}=19$ and $N_{\rm DR3}=32$, yielding $\rho=0.77$. Note that this correlation assumes uniform temporal sampling\citep{Thompson2025AJ}; the actual observations, however, are somewhat uneven and may differ from the GOST predictions.
We repeated the fit and found a marginally lower log-likelihood, but the orbital solution remained consistent with our original result (Figure~\ref{fig:correction}). This indicates that treating the two catalogs as independent is reasonable—at least for the HD\,164604 system.

\bibliography{sample701}{}
\bibliographystyle{aasjournalv7}



\end{document}